\documentclass[a4paper,11pt]{article}

\usepackage{jcappub}
\usepackage{hyperref}
\usepackage{graphicx}
\usepackage{epsfig}
\usepackage[utf8]{inputenc}
\usepackage[export]{adjustbox}
\usepackage{multirow}
\usepackage{amsmath, amssymb, amsfonts,bm}
\usepackage{siunitx}
\usepackage{booktabs}
\usepackage{mathtools}
\usepackage{multirow}
\usepackage{placeins}
\usepackage{stackengine}
\usepackage{fontawesome} 

\newcommand\xrowht[2][0]{\addstackgap[.5\dimexpr#2\relax]{\vphantom{#1}}}
\usepackage[bottom]{footmisc}

\usepackage[
    top    = 2.75cm,
    bottom = 2.60cm,
    left   = 2.50cm,
    right  = 2.50cm]{geometry}

\def\d{\partial}

\def\ln{{\rm ln}}

\def\be{\begin{equation}}
\def\ee{\end{equation}}
\def\bea{\begin{eqnarray}}
\def\eea{\end{eqnarray}}
\def\ba{\begin{align}}

\def\bi{\begin{itemize}}
\def\ei{\end{itemize}}

\def\bx{{\bf x}}
\def\bk{{\bf k}}
\def\bq{{\bf q}}

\def\cG{{\mathcal G}}
\def\cI{{\mathcal I}}
\def\cF{{\mathcal F}}

\title{\fontsize{20}{32}\selectfont{Precision Tests of CO and [CII] Power Spectra Models against Simulated Intensity Maps \vspace{-0.2in} \\}}  

\author[a]{\fontsize{15.}{25}\selectfont Azadeh Moradinezhad Dizgah,}
\author[b]{Farnik Nikakhtar,}
\author[c]{Garrett K. Keating,}
\author[d]{Emanuele Castorina}
\author[]{\vspace{-0.18in} \\}

\affiliation[a]{D\'epartement de Physique Th\'eorique,
Universit\'e de Gen\`eve, 24 quai Ernest Ansermet,\\ 1211 Gen\`eve 4, Switzerland}
\affiliation[b]{Department of Physics and Astronomy, University of Pennsylvania, 209 S. 33rd St., \\ Philadelphia, PA 19104, USA}
\affiliation[c]{Center for Astrophysics, Harvard \& Smithsonian, 60 Garden Street, \\ Cambridge, MA 02138, USA}
\affiliation[d]{Dipartimento di Fisica ‘Aldo Pontremoli’, Universita’ degli Studi di Milano, Via Celoria 16, 20133 Milan, Italy}
\emailAdd{\textcolor{blue}{Azadeh.MoradinezhadDizgah@unige.ch,  farnik@sas.upenn.edu, garrett.keating@cfa.harvard.edu, emanuele.castorina@unimi.it}}

\abstract{Line intensity mapping (LIM) is an emerging technique with a unique potential to probe a wide range of scales and redshifts. Realizing the full potential of LIM, however, relies on accurate modeling of the signal. We introduce an extended halo model for the power spectrum of intensity fluctuations of CO rotational lines and [CII] fine transition line in real space, modeling nonlinearities in matter fluctuations and biasing relation between the line intensity fluctuations and the underlying dark matter distribution. We also compute the stochastic contributions beyond the Poisson approximation using the halo model framework. To establish the accuracy of the model, we create the first cosmological-scale simulations of CO and [CII] intensity maps, \textsf{MithraLIMSims}, at redshifts $0.5 \leq z\leq6$, using halo catalogs from Hidden-Valley simulations, and painting halos according to mass-redshift-luminosity relations for each line. We show that at $z=1$ on scales $k_{\rm max} \lesssim 0.8 \ {\rm Mpc}^{-1}h$, the model predictions of clustering power (with only two free parameters) are in agreement with the measured power spectrum at better than 5\%. At higher redshift of $z=4.5$, this remarkable agreement extends to smaller scale of $ k_{\rm max} \lesssim 2 \ {\rm Mpc}^{-1}h$. Furthermore, we show that on large scales, the stochastic contributions to CO and CII power spectra are non-Poissonian, with amplitudes reproduced reasonably well by the halo model prescription. Lastly, we assess the performance of the theoretical model of the baryon acoustic oscillations (BAO) and show that hypothetical LIM surveys probing CO lines at $z=1$, that can be deployed within this decade, will be able to make a high significance measurement of the BAO. On a longer time scale, a space-based mission probing [CII] line can uniquely measure the BAO on a wide range of redshifts at an unprecedented precision.}

\begin{document}

\maketitle

\section{Introduction}

The large-scale structure (LSS) contains invaluable information about the origin, composition, and evolution of the Universe and provides an incredible opportunity for precision tests of fundamental physics. In addition to ever-increasing precision and volume of upcoming galaxy surveys, such as DESI \cite{Aghamousa:2016zmz}, EUCLID  \cite{Amendola:2016saw}, SPHEREx \cite{Dore:2014cca}, and LSST \cite{Abell:2009aa}, line intensity mapping is emerging as a powerful technique to map a significant fraction of the sky and over extended redshift epochs, largely inaccessible to other LSS probes \cite{Kovetz:2017agg,Kovetz:2019uss,Silva:2019hsh}. In contrast to galaxy surveys, which trace the LSS by resolving individual sources, LIM measures the integrated emission from spectral lines, originating from diffuse IGM and unresolved galaxies (including the faintest ones).  Measurement of spatial fluctuations in the mean line intensity together with the observed frequency of the line provides a 3-dimensional map of the structure, which can be uniquely applied to study cosmology and astrophysics.

Aside from the 21-cm hyper-fine transition of neutral hydrogen, there are several spectral lines of interest for intensity mapping, including the rotational lines of carbon monoxide (CO) \cite{Lidz:2011dx,Breysse:2014uia,Li:2015gqa}, the fine structure line of ionized carbon ([CII]) \cite{Silva:2014ira,Pullen:2017ogs}, as well as H$\alpha$, H$\beta$, Lyman-$\alpha$ and ionized oxygen line species \cite{Silva:2012mtb,Pullen:2013dir,Silva_2017,Gong:2020lim,Yang2021}. Several recent works highlighted the potential of LIM for cosmology, including forecasts of its constraining power for probing nature of dark matter \cite{Creque-Sarbinowski:2018ebl,Bernal:2020lkd} and dark energy \cite{Karkare:2018sar}, primordial non-Gaussianity \cite{MoradinezhadDizgah:2018zrs,MoradinezhadDizgah:2018lac,Liu:2020izx,Chen:2021ykb}, and properties of neutrinos \cite{Gong:2020lim, Bernal:2021ylz, MoradinezhadDizgah:2021upg}. Tentative recent detections of CO intensity shot power spectrum by COPSS \cite{Keating:2015qva,Keating:2016pka,Keenan2021} and mmIME \cite{Keating:2020wlx} surveys, and [CII] emission via cross-correlation between Planck and SDSS quasars \cite{Pullen:2017ogs} have further amplified the growing observational and theoretical interest in LIM. Looking ahead, in addition to several 21-cm surveys, a wide range of ground- and space-based experiments (e.g., COMAP \cite{Li:2015gqa}, TIME \cite{doi:10.1117/12.2057207}, CONCERTO \cite{Lagache_2018}, CCAT-prime \cite{Aravena:2019tye} EXCLAIM \cite{Essinger_Hileman_2020}, SPHEREx \cite{Dore:2014cca} and CDIM \cite{cooray2019cdim}) will provide higher-fidelity detection of line intensity fluctuations at different redshifts from several emission lines \footnote{See \url{https://lambda.gsfc.nasa.gov/product/expt/lim_experiments.cfm} for a list of current and upcoming LIM surveys.}. Success of these surveys will pave the way for next generation wide-field LIM surveys (e.g., \cite{Delabrouille:2019thj,Karkare:2021ryi}), capable of providing precision cosmological constraints.  

Beyond observational challenges (e.g., control of systematics and cleaning of foregrounds \cite{Lidz:2016lub,Cheng:2016yvu,Sun_2018,Cheng:2019,Cheng:2020asz}), unlocking the full potential of LIM relies on accurate modeling of the signal of interest. Similar to traditional galaxy redshift surveys, this requires both state of the art numerical simulations and analytical tools, the latter usually validated by the former. In this work, we present progress in both of these areas, with a new set of LIM mock catalogs, \textsf{Mithra LIMSims}, aimed at cosmological observations, and an improved theoretical model of the line signal. Our mock data of both CO and [CII] lines is generated using the Halo Occupation Distribution (HOD) framework, and it allows for maximal flexibility, which is a desirable feature given the present astrophysical uncertainties of the properties of these emission lines, in particular at high redshifts. We use the halo catalogues from the Hidden Valley (HV) simulations suite \cite{Modi:2019ewx} \href{http://cyril.astro.berkeley.edu/HiddenValley}{\faGlobe}   \footnote{\url{http://cyril.astro.berkeley.edu/HiddenValley}}, which have the volume ($V \simeq 1 \ h^{-3}{\rm Gpc}^3 $) and halo mass resolution ($M_{\rm min} \simeq 10^8 M_\odot$) needed for intensity mapping studies. Therefore, the generated \textsf{Mithra LIMSims} mocks allow very precise measurements of the cosmological observables, required for both forecasting and analysis of data from the next generation of LIM instruments. The presented mocks are complementary to the existing detailed, smaller volume LIM simulations, based on hydrodynamical cosmological simulations and semi-analytic models of galaxy formation (e.g., \cite{Leung2020,Murmu:2021ljb,Yang:2020lpg}). 

The commonly used framework to describe the signal of spectral lines emitted from within galaxies, based on the halo model \cite{Seljak:2000gq,Cooray:2002dia}, relates the intensity of the emission lines to the abundance of halos hosting line-luminous galaxies. It consists of two main ingredients; the relation between line luminosity and halo masses as a function of redshift and the relation between halo properties and the underlying dark matter distribution. Assuming the commonly used empirical semi-analytic relations between the line luminosities and halo mass (see. e.g., \cite{Li:2015gqa}) \footnote{Examples of other parameterization of line luminosity - halo mass relations can be found in Ref. \cite{Padmanabhan:2017ate}.}, in this paper, we study two aspects of the modeling of line power spectrum signal; the impact of nonlinearities in the matter fluctuations and in the line-matter biasing relation, and the deviations from the commonly assumed Poisson shot noise. The former aspect is relevant since although the matter fluctuations are more linear at high redshifts, the nonlinear biasing becomes more critical as the value of the halo/line bias increases with redshifts \cite{Wilson:2019brt}. Therefore, to extract cosmological information from LIM at low and high redshifts, including both nonlinearities is essential. The relevance of the second aspect of the modeling lies in the fact that although on scales smaller than the typical size of the halos, the shot noise is expected to be Poissonian, on larger scales, this is not the case. Inaccurate modeling of these two aspects limits the potential of LIM in constraining both cosmology and astrophysics (e.g., star formation history).    

Our analytic model provides an extension to the standard halo model. Using the EFTofLSS framework \cite{Baumann:2010tm, Carrasco:2012cv, Senatore:2014via, Assassi:2014fva, Assassi:2014fva, Angulo:2015eqa, Senatore:2017pbn}, we include --in the 2-halo term-- the one-loop corrections to line power spectrum, and the Infrared resummation \cite{Senatore:2014via,Senatore:2017pbn,Blas:2016sfa} to account for the smoothing of the Baryon Acoustic Oscillation (BAO) due to large bulk velocities (see. \cite{Smith:2006ne,Valageas:2011up, Schmidt:2015gwz,Philcox:2020rpe} for earlier works). Furthermore, we compute the corrections to line Poisson shot noise on large scales, originating from halo exclusion and clustering on small scales using the halo model \cite{Hamaus:2010im,Baldauf:2013hka,Ginzburg:2017mgf}. We limit our analysis to real space (neglecting the redshift-space distortions) and consider only the auto-correlations of the CO and [CII] lines, leaving a multi-line study \cite{Sun2019,Schaan:2021gzb, Schaan:2021hhy,Yang:2020lpg} to future works. While we consider only CO and [CII] lines, the extended halo model we construct applies to other lines emitted from galaxies. Similarly, our implemented prescriptions for producing simulated intensity maps are straightforwardly extendable to other lines originating from within galaxies. 

Together with this paper, we release an accompanying software package, \textsf{limHaloPT} \href{https://github.com/amoradinejad/limHaloPT.git} {\faGithub} \footnote{\url{https://github.com/amoradinejad/limHaloPT}} (written in C), which can be used to compute the theoretical model of the line signal presented here. The code takes the linear matter power spectrum from CLASS \cite{Blas2011} \href{https://github.com/lesgourg/class_public} {\faGithub} \footnote{\url{https://github.com/lesgourg/class_public}} as an input. Furthermore, we make public an extension \href{https://github.com/farnikn/MithraLIMSims}{\faGithub} \footnote{\url{https://github.com/farnikn/MithraLIMSims}} of the analysis pipeline of HV simulations \href{https://github.com/modichirag/HiddenValleySims} {\faGithub} \footnote{\url{https://github.com/modichirag/HiddenValleySims}} , which includes the HOD prescriptions for painting the halos with line intensity. The sub-sampled grids of a select CO and [CII] intensity maps used in this paper will be soon made public \href{http://cyril.astro.berkeley.edu/MithraLIMSims}{\faGlobe}  \footnote{\url{http://cyril.astro.berkeley.edu/MithraLIMSims}}. 

The rest of the paper is organized as follows: in Section \ref{sec:theory} we present the extended halo model of line intensity power spectrum. In Section \ref{sec:sim_th}, after describing the specification of the numerical simulations, we present the comparison of the theoretical predictions against the measurements of clustering and shot powers and present the prospects of the measurement of the BAO signal by future LIM surveys. Finally in section \ref{sec:summary}, we draw our conclusions. Additional details of the calculations are presented in two Appendices.

\section{The Halo Model of Line Power Spectrum}\label{sec:theory}

The total line intensity power spectrum consists of clustering and shot-noise contributions. Within the halo-model framework, the clustering component can be further split into 1- and 2-halo terms, the former capturing the contribution to the fluctuations from line emitters residing within the same halo, and the latter accounting for clustering of line emitters in two different halos. On scales smaller than the typical sizes of halos, the shot-noise approaches the prediction of Poisson statistics. On large scales, however, deviations from Poisson shot noise are expected and can be computed using the halo model. In the rest of this section, we carry out the computation of each of these contributions, i.e.,
\be\label{eq:HM}
P_{\rm line}^{\rm tot}(k, z) = P_{\rm line}^{1h}(k,z)+ P_{\rm line}^{2h}(k,z) + P_{\rm line}^{\rm shot}(k,z).
\ee

Before delving into the detailed description of each term in Eq.  \eqref{eq:HM}, we note that despite being a powerful framework to describe the clustering statistics of matter and biased tracers, the analytical halo model has several shortcomings, including not enforcing the conservation of matter stress-energy, not recovering the perturbation theory prediction on large-scales, not having a consistent description of 2- and 3-point statistics, and not describing the transition regime of 1- and 2-halo terms accurately. We neglect these limitations here and refer the interested reader to several previous works that address these issues \cite{Valageas:2011up,Valageas:2010yw,Mohammed:2014lja,Seljak:2015rea,Schmidt:2015gwz}. 

\subsection{Clustering Power}\label{sec:clust}

Assuming that the line intensity continuously traces the dark matter halos according to an NFW profile and is not associated with galaxies, i.e., there is no splitting of the galaxies to central and satellites \footnote{See Ref. \cite{Wolz:2018svc} for an extensive discussion of HI models with and without the central/satellite distinction.}, the 1- and 2-halo terms are given by 
\be\label{eq:1h}
P_{\rm line}^{1h}(k,z) = \left(\frac{c^4 p_{2,\sigma}}{4k_B^2 \nu_{\rm obs}^4 } \right)\int_{M_{\rm min}}^{M_{\rm max}} dM \ n(M,z) u(k|M)^2{\left[\frac{L(M,z)}{4 \pi \mathcal D_L^2} 
\left ( \frac{dl}{d\theta} \right )^{2} \frac{dl}{d\nu} \right ]}^2, 
\ee
\begin{align}\label{eq:2h}
P_{\rm line}^{2h}(k, z) &= \left(\frac{c^4 p_{1,\sigma}^2
}{4k_B^2 \nu_{\rm obs}^4 }\right) \int_{M_{\rm min}}^{M_{\rm max}} dM_1  \ n(M_1,z) u(k|M_1) {\left[\frac{L(M_1,z)}{4 \pi \mathcal D_L^2} \left ( \frac{dl}{d\theta} \right )^{2} \frac{dl}{d\nu} \right ]}\nonumber \\
 &\times  \int_{M_{\rm min}}^{M_{\rm max}} dM_2  \ n(M_2,z) u(k|M_2) {\left[\frac{L(M_2,z)}{4 \pi \mathcal D_L^2} \left ( \frac{dl}{d\theta} \right )^{2} \frac{dl}{d\nu} \right ]} \ P_h^{M_1,M_2}(k,z),
\end{align}
where $n(M,z)$ is the halo mass function, which we take to be the Sheth-Tormen function (ST) \cite{Sheth:1999mn}, $u(k|M)$ is the Fourier transform of halo profile, and ${\mathcal D}_L$ is the luminosity distance. $L(M,z)$ is the specific luminosity of the galaxy located in a halo of mass $M$ at redshift $z$ hosting galaxies luminous in a given line. The terms $dl/d\theta$ and $dl/d\nu$ reflect the conversion from units of comoving lengths, $l$, to those of the observed specific intensity: frequency, $\nu$, and angular size, $\theta$. The term $dl/d\theta$ is equivalent to comoving angular diameter distance, whereas $dl/d\nu = c(1+z)/[\nu_{\rm obs}H(z)]$, with $H(z)$ being the Hubble parameter at a given redshift.  

In Eqs. \eqref{eq:1h} and \eqref{eq:2h}, the parameters $p_{n,\sigma}$ account for the scatter in the relation of line luminosity and halo masses. Assuming a log-normal scatter, we have
\be\label{eq:scatter}
p_{n,\sigma} = \int_{-\infty}^\infty dx \frac{10^{n x}}{\sqrt{2\pi}\sigma_{\rm line}} e^{-x^2/2\sigma_{\rm line}^2}.
\ee
This scatter originates from fluctuations in various astrophysical process on which the line luminosity depends on, such as star formation, molecular growth, metal enrichment, and different phases of interstellar medium, and has been reported in detailed hydrodynamical simulations, e.g.,  in Refs. \cite{Lagache_2018,Leung2020} for [CII]. While in principle there could be several independent scatter parameters to capture the relevant physical process (see e.g. Ref. \cite{Li:2015gqa}), as in Ref. \cite{Keating:2016pka} we assume a single scatter parameter in the relation between line luminosity and star formation rate (SFR). 

We set the value of  $\sigma_{\rm line} = 0.37$, corresponding to the fiducial model of Ref. \cite{Li:2015gqa}, which is in reasonably good agreement with observations of Refs. \citep{Sargsyan2012,Speagle2014,Carilli:2013qm,Kamenetzky_2016}. This scatter, results in enhancements of both 1- and 2-halo contributions of clustering power spectrum, affecting the 1-halo term (proportional to the second moment of the line luminosity) more significantly since $p_{2,\sigma} > p_{1,\sigma}^2$. Our choice of $\sigma_{\rm line}$ for the log-scatter, enhances the 1- and 2-halo terms by factors of $\sim2$ and $\sim4$, respectively. Worthy to note that the recent results of Ref. \cite{Murmu:2021ljb}, in which the scatter inferred from the hydrodynamic and radiative transfer simulation of early galaxies \cite{Leung2020} was applied to the halo distribution of a large-volume Nbody simulations, found an enhancement of the [CII] power spectrum by factor of $\sim3$ for [CII] power spectrum at $z=6$.

\subsubsection{Line Luminosities}

We model the luminosities of CO rotational ladder by assuming scaling relations between CO and far-infrared (FIR) luminosities and between FIR luminosity and SFR, following Li {\it et al.} \cite{Li:2015gqa}. We use the empirical fit of Behroozi {\it et al.} \cite{Behroozi:2012sp, Behroozi:2012iw} to relate the SFR and the halo mass and redshift, and use the Kennicut scaling relation \citep{Kennicutt:1998zb} to relate the SFR to FIR luminosity,
\be \label{eq:SFR_IR}
{\rm SFR}(M,z) = \delta_{\rm MF} \times 10^{-10} L_{\rm IR}. \
\ee
The normalization $\delta_{\rm MF}$ depends on the assumptions of initial mass function, the duration of star-formation, etc. As in Refs. \cite{Behroozi:2012sp, Li:2015gqa}, we take $\delta_{\rm MF} =1 \ M_\odot {\rm yr}^{-1} L_\odot^{-1}$. We use the empirical fit of Kamenetzky {\it et al.} to Herschel SPIRE Fourier Transform Spectrometer data \cite{Kamenetzky_2016}, to relate the CO and FIR luminosities as a log power-law 
\be\label{eq:CO_lum_IR}
\log L_{\rm IR} = \alpha \log  L'_{\rm CO} + \beta,
\ee
where the CO line luminosity is in unit of ${\rm K} \ {\rm km} \ s^{-1} {\rm pc}^2$, and $L_{\rm IR}$ is in unit of $L_\odot$. Values of $\alpha$ and $\beta$ for the three CO lines we consider in this paper are given in Table \ref{tab:CO_models}. The CO rotational ladder luminosities (in unit of $L_\odot$) are then given by
\be
L_{\rm CO(J\rightarrow J-1)}[L_\odot] = 4.9 \times 10^{-5} J^3 \ L'_{\rm CO(J\rightarrow J-1)}
\ee

\begin{table}[t]
\par\smallskip
\centering
\begin{tabular}{c c c  }
\hline \hline
${\rm model}$ 	& $ \alpha$ & $\beta$  \\  \hline 
CO(2-1) & 1.11 & 0.6  \\
CO(3-2) & 1.18 &  0.1   \\
CO(4-3) & 1.09 & 1.2 \\
\hline
\end{tabular}
\caption{Parameters of the models of CO luminosities from empirical fits of Ref. \cite{Kamenetzky_2016}}
\label{tab:CO_models} 
\end{table} 

For [CII] fine transition line, we use model ${\bf m}_1$ of Silva {\it et al.} \cite{Silva:2014ira}, which relates the [CII] luminosity and SFR as a log power-law,
\be \label{eq:CII_lum}
\log L_{\rm [CII]} = a \times \log {\rm SFR}(M,z) +b,
\ee 
with $a = 0.8475, \ b = 7.2203 $. This model corresponds to the fit to high redshift galaxies by De Looze {\it et al} \cite{DeLooze:2014dta}. While in Ref. \cite{Silva:2014ira}, the SFR and halo mass were related using post-processed simulated galaxy catalogs presented in Refs.  \cite{DeLucia:2006szx, Guo2011} from the Millennium simulations \cite{Springel:2005nw}, here we use the empirical fit by Behroozi {\rm et al} for ${\rm SFR}(M,z)$ relation. 

\subsubsection{Halo Profile}

For the results shown in this section, we will use the NFW halo profile, truncated at virial radius $r_{\rm vir}$, which in real-space is given by
\be
u(r|M) = \frac{\rho_s}{(r/r_s)(1+r/r_s)^2},
\ee
where the radius $r_s$ is related to the virial radius via $r_s \equiv r_{\rm vir}/c$ for halo concentration $c(M,z)$. We neglect the possible dependence of the concentration on the underlying density field or tidal tensor \cite{Schmidt:2015gwz,Schmidt:2018hbj} and use the parameterization from Duffy {\it et al.} \cite{Duffy:2008pz}
\be
c(M,z) = A\left(\frac{M}{10^{12} h^{-1}M_\odot } \right)^\alpha (1+z)^\beta,
\ee
with $A= 7.85, \alpha = -0.081, \beta=-0.71$. The normalization factor $\rho_s$ is given by
\be
\rho_s = \frac{M}{4\pi r_s^2[\log(1+c)-\frac{c}{1+c}]}.
\ee
In the comparison of the model with simulations, presented in the proceeding sections, we will only consider the 2-halo term on relatively large scales, where the halo profile assymptotes to unity, $u(k|M) \rightarrow 1$.

\subsubsection{Nonlinearities}

In the expression of 2-halo term, $P_h^{M_1,M_2}$ is the power spectrum of two halos with masses $M_1$ and $M_2$. We first review the ingredients for modeling the halo power spectrum at one-loop in perturbation theory. The reader familiar with the details can skip to the second half of this section, where we present the halo model prediction of the line power spectrum, including nonlinear corrections.

We consider the halo bias expansion in terms of renormalized operators \cite{McDonald:2006mx,Assassi:2014fva} constructed from density field, tidal shear, as well as non-local higher derivative operators of density and gravitational potential. We drop the stochastic terms in the bias expansion, and instead compute the shot noise using halo model (see Section \ref{sec:th_shot}). Assuming Gaussian initial conditions, keeping terms up to third order and neglecting the stochastic contributions, the halo bias expansion takes the form (see Ref. \cite{Desjacques:2016bnm} for an extended review on the topic),
\begin{align}\label{eq:bias}
\delta_h(\bx,z)&=b_1\delta(\bx,z)+b_{\nabla^2\delta}\nabla^2\delta(\bx,z)+ \frac{b_2}{2} \delta^2(\bx,z)+ b_{\mathcal{G}_2}\mathcal{G}_2(\bx,z)  \nonumber \\
&+ \frac{b_3}{6} \delta^3(\bx,z) + b_{\mathcal{G}_3}\mathcal{G}_3(\bx,z) + b_{(\mathcal{G}_2\delta)}\mathcal{G}_2(\bx,z)\delta(\bx,z) +b_{\Gamma_3}\Gamma_3(\bx,z), 
\end{align}
where $\delta$ is the matter over density field, $\cG_2$ and $\cG_3$ are the second and third order Galileon operators  \cite{Chan:2012jj}
\begin{align}
\mathcal{G}_2(\Phi) & \equiv(\d_i\d_j\Phi)^2-(\d^2\Phi)^2, \\
\mathcal{G}_3(\Phi) & \equiv-\d_i\d_j\Phi\d_j\d_k\Phi\d_k\d_i\Phi-\frac{1}{2}(\d^2\Phi)^3 +\frac{3}{2}(\d_i\d_j\Phi)^2\d^2\Phi\,, 
\end{align}
while $\Gamma_3$ is the difference between density and velocity tidal tensors,
\begin{align}
\Gamma_3 &\equiv\mathcal{G}_2(\Phi)-\mathcal{G}_2(\Phi_v),
\end{align}
where $\Phi$ and $\phi_v$ are the density and velocity potentials. All the bias parameter depend on halo mass and redshift, which we have dropped for brevity. Furthermore, while not explicitly indicated in our notation, all quadratic and cubic in fluctuation fields (i.e. composite operators) are renormalized \cite{Assassi:2014fva}.

\begin{figure}[t]
    \centering
    \includegraphics[width=0.49\textwidth]{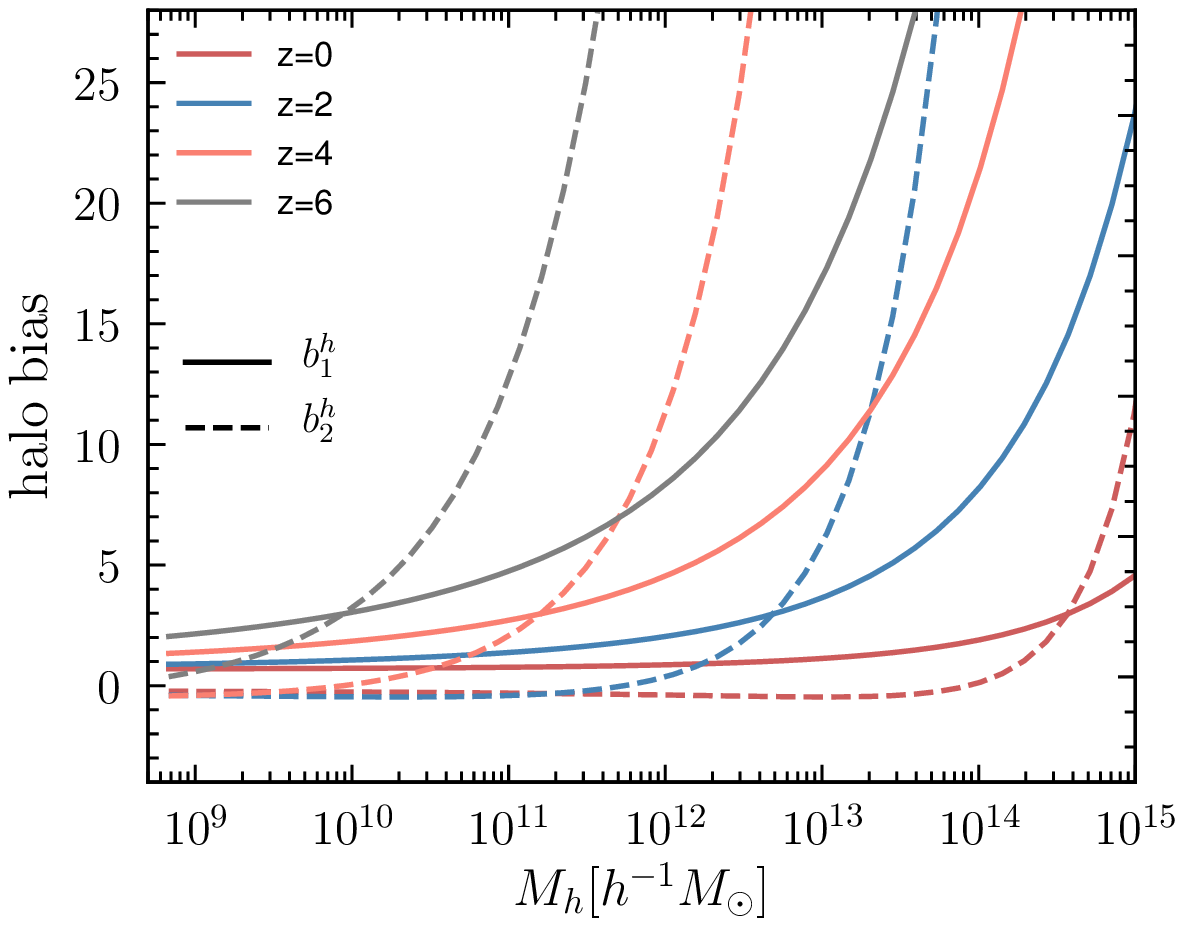}
    \includegraphics[width=0.49\textwidth]{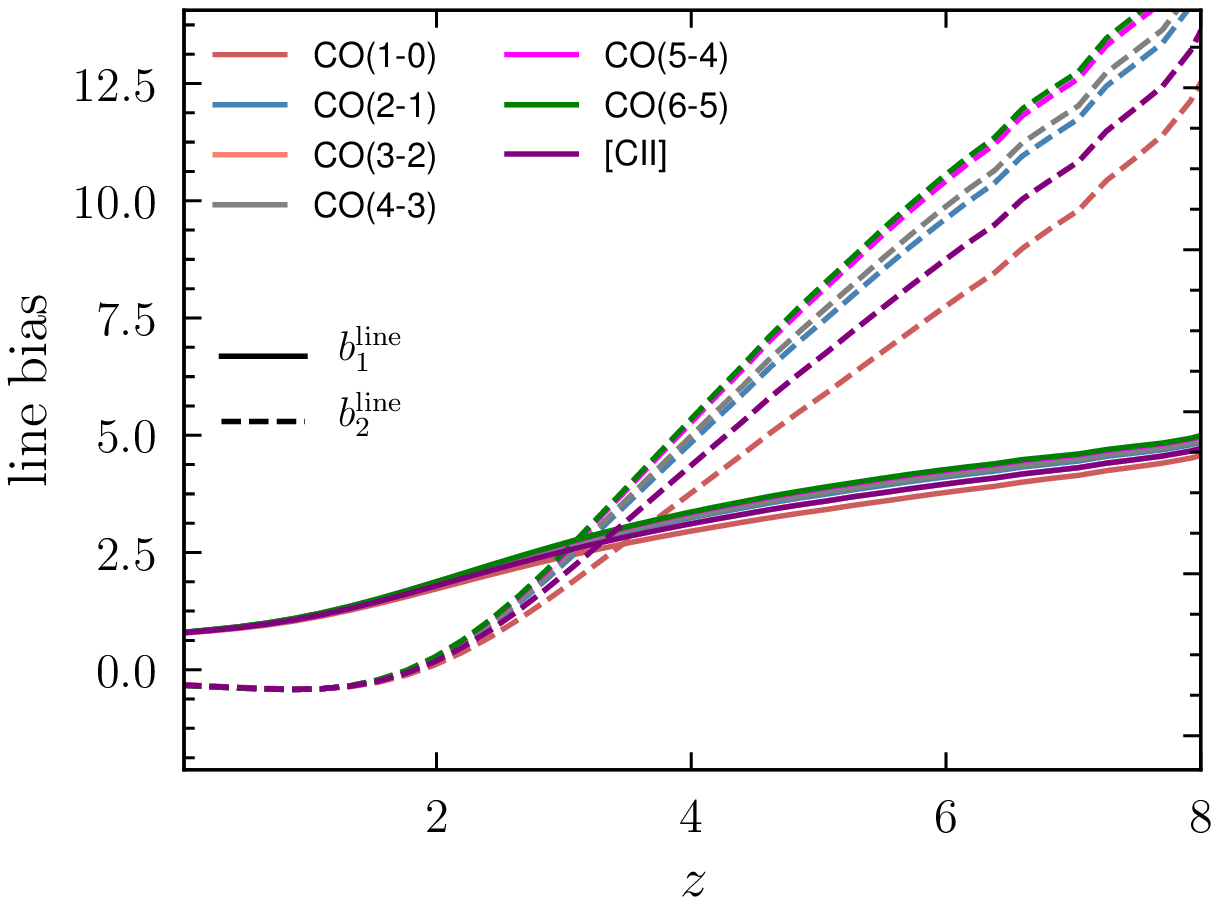}\vspace{-.1in}   
    \caption{{\it Left}: ST linear (solid lines) and quadratic (dashed lines) halo biases at several redshifts. {\it Right}: linear and quadratic line biases for CO rotational ladder (up to J=6) and [CII] lines given in Eq.  \eqref{eq:line_bias}.}
    \label{fig:th_bias}
\end{figure}

For two halos with the same mass, $M_1 = M_2$, the one-loop halo power spectrum is given by \cite{Assassi:2014fva},
\begin{align}\label{eq:ph_1loop}
P_h(k,z)&= b_1^2 \tilde b_{\nabla^2} k^2 P_0(k,z) + b_1^2 P_m^{1{\rm loop}}(k,z) + b_1b_2\mathcal{I}_{\delta^2}(k,z)+2b_1b_{\mathcal{G}_2}\mathcal{I}_{\mathcal{G}_2}(k,z) \notag \\
&+\frac{1}{4} b^2_2\mathcal{I}_{\delta^2\delta^2}(k,z)+b^2_{\mathcal{G}_2}\
\mathcal{I}_{\mathcal{G}_2\mathcal{G}_2}(k,z) +b_2b_{\mathcal{G}_2}\mathcal{I}_{\delta^2\mathcal{G}_2}(k,z) \notag \\ &+2b_1(b_{\mathcal{G}_2}+\frac{2}{5}b_{\Gamma_3})\mathcal{F}_{\mathcal{G}_2}(k,z),
\end{align}
where $P_0$ is the linear matter power spectrum. The explicit expressions for the loop integrals $\mathcal{I}_{\mathcal{O}}$ and $\mathcal{F}_{\mathcal{O}}$ are given in Appendix \ref{app:1loop}. In the analytical results that we show in this paper, we use the ST predictions of linear and quadratic local-in-matter halo biases, $b_1, b_2$ \cite{Sheth:1999mn, Scoccimarro:2000gm} (shown in the left panel of Figure \ref{fig:th_bias}), and use the co-evolution prediction for the tidal shear biases, $b_{\mathcal{G}_2}, b_{\Gamma_3}$ (see Appendix \ref{app:halo_biases} for explicit expressions). 

The first term in Eq.  \eqref{eq:ph_1loop} is an effective term that accounts for two contributions;  the linear higher derivative bias operator, $b_{\nabla^2}$, capturing non-locality of halo formation \cite{Desjacques:2010gz}, and the EFT counter term, $c_s^2$, accounting for the impact of the non-vanishing small-scale dark matter stress-tensor on large-scale fluctuations \cite{Carrasco:2012cv}. While these two effects become significant at different scales (the former on scales corresponding to the size of the halos and the latter at the nonlinearity scale), they are indistinguishable from one another in the power spectrum as they have the same $k^2$ dependence \footnote{Note that $c_s^2$ is a positive value and is related to the time dependence of the stress-tensor, while $b_{\nabla^2}$ can be either positive or negative.}. We, therefore, model both contributions with a single effective parameter $\tilde b_{\nabla^2}$ (which can be positive or negative). In evaluating this term, since on small scales $k^2$ become unbounded, following Ref. \cite{Philcox:2020rpe}, we use the Pade approximation, 
\be
b_1^2 \tilde b_{\nabla^2} k^2 P_0(k,z) \ \longrightarrow \  b_1^2 \tilde b_{\nabla^2} \left [ \frac{k^2}{1+(k/\hat k)^2}\right] P_0(k,z).
\ee
This form has the right asymptotic behavior on large scales and is finite on small scales. The exact value of $\hat{k}$ is not of great importance since it is degenerate with $\tilde b_{\nabla^2}$. In our numerical evaluations, we set $\hat k = 2.5 \ h/{\rm Mpc}$, which we found to provide a good fit to the measured power spectra, fitting by eye the $\tilde b_{\nabla^2}$ parameter. 

In addition to including the loop contributions to DM power spectrum, $P_{1\rm loop}$, and EFT counter term, we also model the IR resummation to account for the damping of the BAO due to large-scale relative displacements. In our implementation of the IR resummation, we follow the approach of Ref. \cite{Blas:2016sfa}, without adding any new ingredients. Therefore, we only report the final expressions here. Since the long displacements only affect the BAO wiggles, we split the linear matter power spectrum into smooth $ P_{\text{nw}}$ and wiggly parts $P_{\text{w}}$, 
\be
P_0(k,z)= P_{\text{nw}}(k)+P_{\text{w}}(k,z)\,,
\ee
and apply the IR ressumation to the wiggle part, which at leading order appears as an exponential suppression factor,
\be\label{eq:PO_real}
P_{\text{LO}}(k,z) \equiv P_{\text{nw}}(k,z)+e^{-k^2\Sigma^2}P_{\text{w}}(k,z)\,.
\ee
The damping exponent is given by
\be
\Sigma^2 \equiv\frac{4\pi}{3}\int_0^{k_s}dq \ P_{\text{nw}}(q,z) \left[1-j_0\left(\frac{q}{k_{\rm osc}}\right)+2j_2\left(\frac{q}{k_{\rm osc}}\right)\right]\,.
\ee
Here, $k_{\rm osc}$ is the inverse of the BAO scale $\sim 110\ h^{-1}{\rm Mpc}$, $k_s$ is the separation scale controlling the modes to be resummed \footnote{In principle, $k_s$ is arbitrary and any dependence on it should be treated as a theoretical error.}, and $j_n$ are the spherical Bessel function of order $n$. At next-to-leading order one uses the expression in Eq. \eqref{eq:PO_real} as an input in one-loop power spectrum,
\be
P_{\text{NLO}}(k,z) \equiv P_{\text{nw}}(k,z) +e^{-k^2\Sigma^2}P_{\text{w}}(k,z)(1+k^2\Sigma^2) + P_{\text{1-loop}}[P_{\text{nw}}+e^{-k^2\Sigma^2}P_{\text{w}}](k,z)  \,,
\ee
where $P_{\text{1-loop}}$ should be considered a functional of the leading-order IR-resummed DM power spectrum. Similarly, the linear DM power spectra in all the (halo) bias loops in Eq. \eqref{eq:ph_1loop} are replaced by $P_{\rm LO}$. To  apply  the  IR  resummation,  we  need  to  implement  an  algorithm  to  split  the  power spectrum into wiggle and no-wiggle contributions. Since this splitting is not unique, different methods result in differences in the broadband and wiggles extracted \footnote{See Refs. \cite{Hamann:2010pw,Vlah:2015zda,MoradinezhadDizgah:2020whw} for comparisons of various methods.}. Nevertheless, the impact on next-to-leading order, IR resummed power spectrum are less than 0.3\% on scales $k \leq 0.6 \ h/{\rm Mpc}$ \cite{MoradinezhadDizgah:2020whw}. We use a 1-dimensional Gaussian filter on logarithmic scale to smooth the DM power spectrum and extract the broadband \cite{Vlah:2015zda}. We set the smoothing scale to $\lambda = 0.25 \ {\rm Mpc}^{-1}h$. The no-wiggle power spectrum is then given by
\be\label{eq:pk_rescaled}
P_{\rm nw}(k) = P_{\rm approx}(k) \cF\left[P(k)/P_{\rm approx}(k)\right],
\ee
where 
\be
\cF = \frac{1}{\sqrt{2\pi\lambda}} \int d \ \ln q \ P(q)/P_{\rm approx}(q) \ {\rm exp}\left(- \frac{1}{2\lambda^2}(\ln \ k - \ln \ q)^2\right).
\ee
We use the Eisentein \& Hu fit \cite{Eisenstein:1997ik} for the approximate power spectrum.

Having described the nonlinear model of halo power spectrum, we now return to line power spectrum. In Eq. \eqref{eq:ph_1loop}, the only quantities that depend on the halo mass are the bias parameters. Therefore, when computing the 2-halo contribution to line intensity power spectrum, the mass integration over the one-loop halo power spectrum can be re-written in terms of luminosity weighted line bias parameters. Since the biases are also weighted by the halo profile, they become scale-dependent. Considering the large scales, where $u(k|M) \rightarrow 1$, we have
\be\label{eq:line_bias}
b_{\rm x}^{\rm line}(z) = \frac{\int_{M_{\rm min}}^{M_{\rm max}}d M \ n(M,z) \ b_{\rm x}(M,z) L(M,z)}{\int_{M_{\rm min}}^{M_{\rm max}} d M \ n(M,z)L(M,z) },
\ee
where ``x" refers to any of the biases appearing in Eq.  \eqref{eq:bias}. To illustrate the relative size of linear and quadratic biases, in Fig. \ref{fig:th_bias}, on the left we show the halo biases as a function of halo mass for several redshifts, and on the right the biases of the considered CO rotational lines, and the [CII] line as a function of redshift.

On large-scales, where the halo bias is approximately linear, and the halo profile asymptotes to unity, the expression of the 2-halo term in Eq.  \eqref{eq:2h} reduces to the commonly used tree-level model (which neglects the nonlinearities in matter fluctuations and biasing),
\be\label{eq:pk_tree}
P_{\rm line}^{\rm tree} (k,z) = \left[\bar T_{\rm line}(z) b_1^{\rm line}(z)\right]^2 P_0(k,z), 
\ee
where $\bar T_{\rm line}$ is the mean brightness temperature of the line given by
\be
\bar T_{\rm line}(z) = \left(\frac{c^2 p_{1,\sigma}}{2k_B \nu_{\rm obs}^2 } \right)\int_{M_{\rm min}}^{M_{\rm max}} dM \ n(M,z) \left[\frac{L(M,z)}{4 \pi \mathcal D_L^2} 
\left ( \frac{dl}{d\theta} \right )^{2} \frac{dl}{d\nu} \right]. 
\ee

\begin{figure}[t]
    \centering
    \includegraphics[width=0.49 \textwidth]{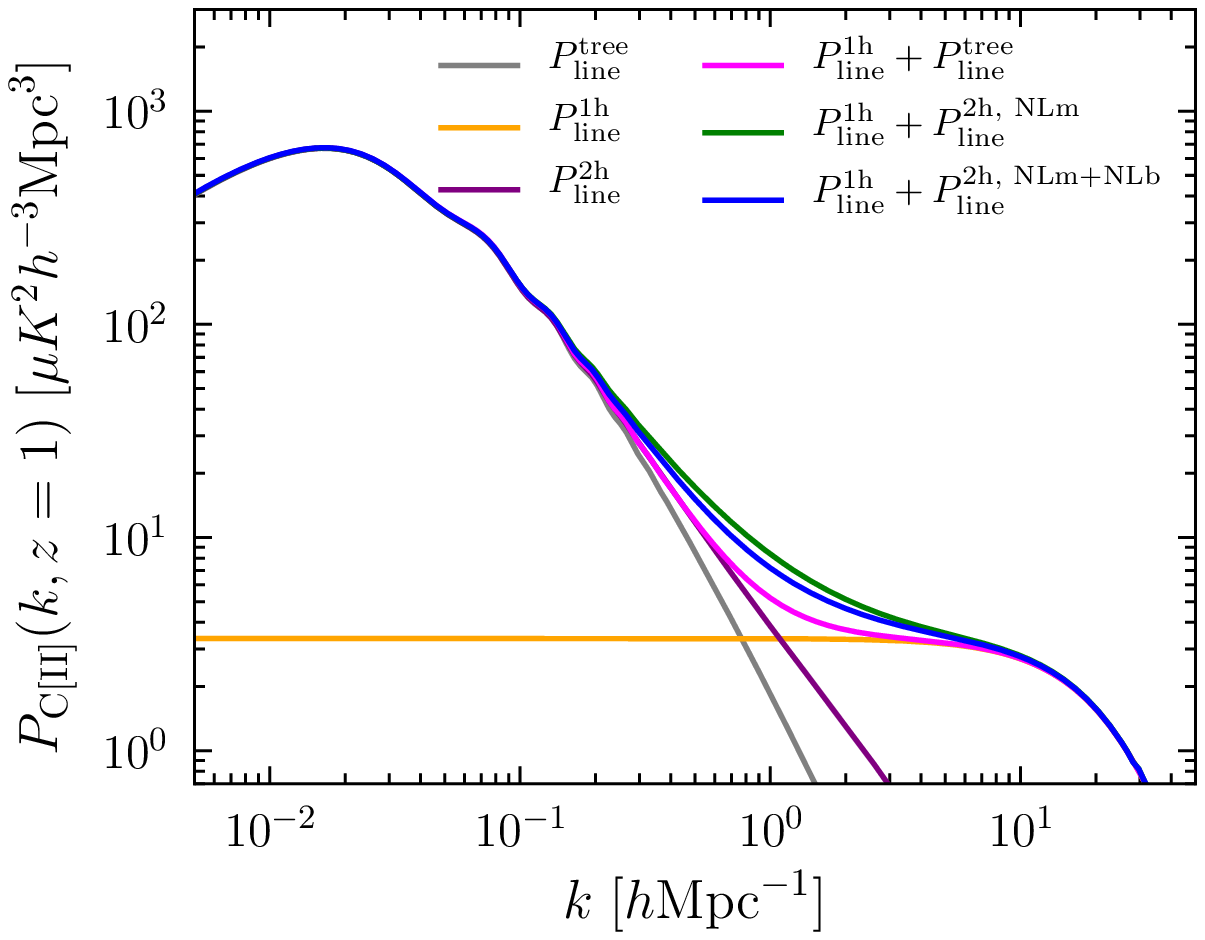}
    \includegraphics[width=0.49 \textwidth]{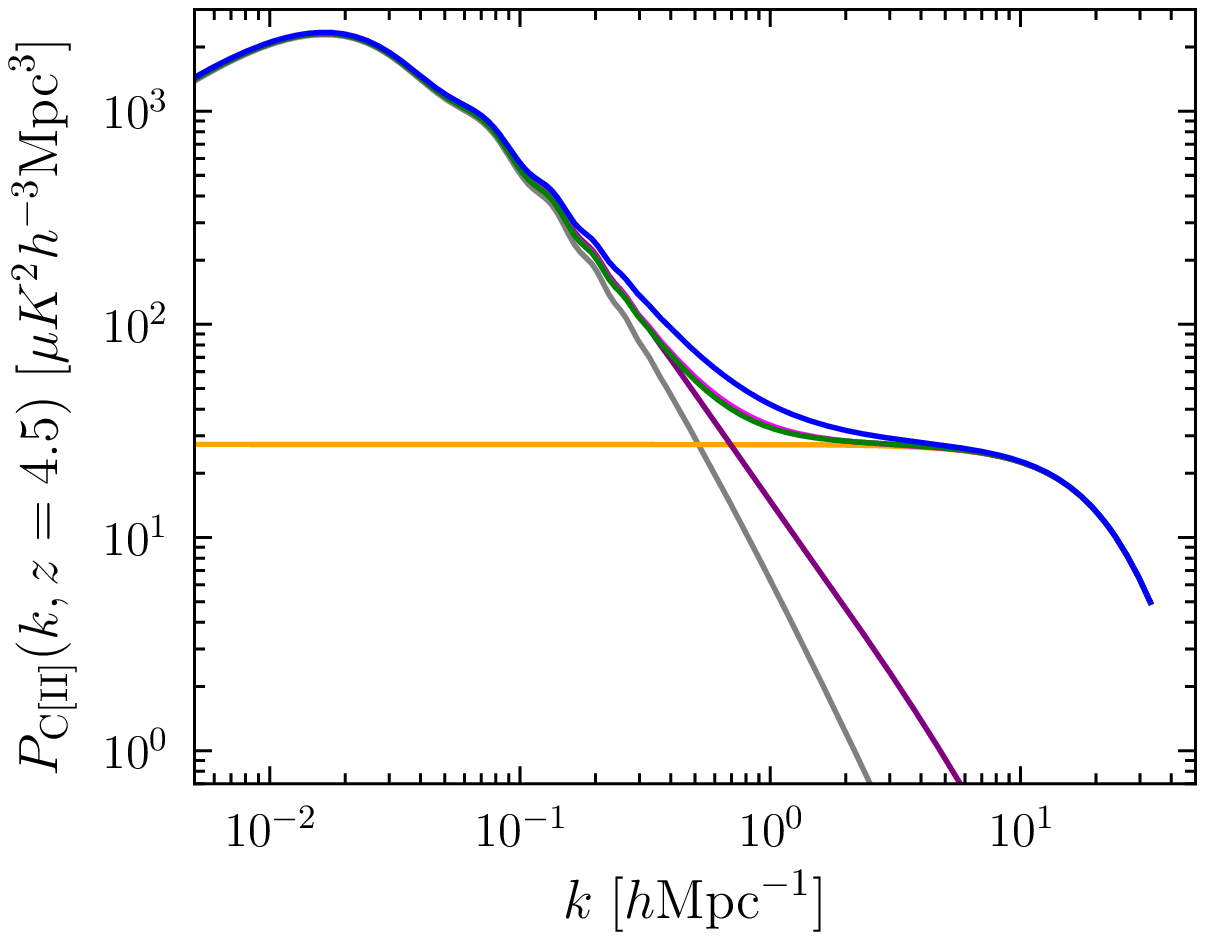}\vspace{-.1in}
    \caption{Halo model predictions for [CII] power spectrum, at $z=1$ on the left and at $z=4.5$ on the right. The gray, yellow and purple lines correspond to Eqs. (\ref{eq:pk_tree}, \ref{eq:1h}, \ref{eq:2h}). Also shown are the sum of 1- and 2-halo term, neglecting all nonlinear corrections (magenta), including nonlinearities of only matter fluctuations (green), and both matter and line biases (blue) in the 2-halo term. In the right plot, the magenta and green lines nearly overlay. We assumed Sheth-Tormen mass function, and NFW profile for [CII] emitting galaxies.} 
    \label{fig:pk_HM}
\end{figure}
In Figure \ref{fig:pk_HM}, we show the theoretical predictions for [CII] power spectrum at $z=1$ (on the left) and $z=4.5$ (on the right), including the tree-level model of Eq. \eqref{eq:pk_tree} (in gray), and the 1- and 2-halo terms (in orange and purple). The purple line, includes all nonlinear terms. To illustrate the importance of accounting for nonlinearities of matter and line biases, we also show the sum of 1- and 2-halo terms, using tree-level 2-halo term (in magenta), including only one-loop contributions to matter (in green), and including both the matter and bias loops (in blue). The overall amplitude of the line power spectrum (both 1- and 2-halo terms) at $z=4.5$ is larger than at $z=1$. We note, however, that the redshift-dependence of the amplitude is not monotonic, but instead has a rise and fall (see Figure 3 of Ref. \cite{MoradinezhadDizgah:2021upg}) and peaks at $z\simeq 2$. This behavior is largely driven by the cosmic star formation history which determines the line luminosity. The relative change of matter power spectrum and linear line bias also affect the redshift-dependence of the 2-halo term. Regarding nonlinear contributions to the line power spectrum, at $z=1$, the loop corrections to matter power spectrum are the dominant corrections, since the nonlinear biases are still small (see Figure \ref{fig:th_bias}). For the values of higher order biases we assumed, including both the one-loop corrections of matter and biases, reduces the total nonlinear corrections compared to matter-only loops. On the other hand, at $z=4.5$, the matter fluctuations are more linear, while nonlinear biases have grown large. Therefore, the bias loops are the dominant nonlinear correction.

\subsection{Shot-Noise Beyond The Poisson Limit} \label{sec:th_shot}

The shot noise of the biased tracers is not strictly Poissonian; halo exclusion \cite{Hamaus:2010im,Baldauf:2013hka,Ginzburg:2017mgf,Schmittfull:2018yuk} and nonlinear halo clustering outside the exclusion radius \cite{Heavens:1998es,Smith:2006ne} violate the assumption of the Poisson model that the tracers are randomly sampled within a given volume. Based on the results of Refs. \cite{Hamaus:2010im,Baldauf:2013hka,Ginzburg:2017mgf}, we use the halo model framework to compute the corrections to the shot-noise of line intensity fluctuations on large scales, which for dark matter halos in N-body simulations is shown to reproduce reasonably well the measured stochasticity on large-scales \cite{Hamaus:2010im}. This agreement is despite some of the unphysical prediction of the analytical halo model, namely a non-zero white noise in the auto and cross spectra of halo and matter density fields. We neglect these issues and refer the interested reader to Refs. \cite{Schmidt:2015gwz,Ginzburg:2017mgf,Chen:2019wik} for some proposed remedies.

The stochasticity in the halo abundance and other biased tracers such as intensity fluctuations of spectral lines does not correlate with matter overdensity on large scales, i.e. with initial conditions. Therefore, one can isolated the stochastic contribution to the power spectrum in simulations by subtracting the part correlated with long-wavelength correlations and defining the stochastic field as \cite{Hamaus:2010im},
\be\label{eq:stoch}
\epsilon_{\rm line}(k,z) = \delta_{\rm line}(k,z) - \bar T_{\rm line}(z) b_{\rm line}(z) \delta(k,z).
\ee
The power spectrum of the line shot-noise (i.e., the noise covariance) defined as 
\be
 P_{\epsilon \epsilon}^{\rm line}(k,z) =  (2\pi)^3 \delta_D({\bf k}+{\bf k'})  \langle\epsilon_{\rm line} ({\bf k},z) \epsilon_{\rm line}({\bf k'},z)\rangle 
\ee
is thus given by 
\begin{align}\label{eq:shot_power}
P_{\epsilon \epsilon}^{\rm line}(k,z) &=  P_{\rm line}(k,z) - 2 \bar T_{\rm line}(z)b_1^{\rm line}(z)P_{\rm line,\delta}(k,z) + \left[\bar T_{\rm line}(z) b_1^{\rm line}(z)\right]^2 P_{\rm \delta \delta}(k,z). 
\end{align}

On the large-scales, where $ u(k|M) \rightarrow 1$, the 2-halo contributions to the auto and cross spectra of line and matter are given by 
\begin{align}\label{eq:2h_ls}
   &\lim_{k\rightarrow0}  P_{\rm line }^{2h}(k,z) \hspace{.08in} = \left[\bar T_{\rm line} b_1^{\rm line}(z)\right]^2 P_0(k,z) + \frac{1}{4} \left[ \bar T_{\rm line} b_2^{\rm line}(z)\right]^2 \int_{\bf q} P_0^2(q,z), \notag \\
   &\lim_{k\rightarrow0}  P_{\rm line,\delta}^{2h}(k,z) = \bar T_{\rm line}b_1^{\rm line}(z) P_0(k,z), \notag \\
   &\lim_{k\rightarrow0}  P_{\delta\delta}^{2h}(k,z) \hspace{.12in} = P_0(k,z). 
\end{align}
In the first line, $\int_{\bf q} \equiv d^3 q$ is the 3-dimensional Fourier integration. The term proportional to quadratic bias $b_2$ in the first line of Eq.  \eqref{eq:2h_ls} is the low-k limit of the loop contribution corresponding to ${\mathcal I}_{\delta^2 \delta^2}$ term in Eq.  \eqref{eq:ph_1loop}, which appears as a constant, and hence should be absorbed as a shot-noise \cite{McDonald:2006mx}. This is the only surviving 2-halo term contributing to the shot noise, as the rest cancel one another. 

Therefore, the shot-noise power spectrum reduces to \begin{align}\label{eq:ls_shot_power}
\lim_{k\rightarrow0} P_{\epsilon \epsilon}^{\rm line}(k,z) &=  \frac{1}{4} \left[ \bar T_{\rm line} b_2^{\rm line}(z)\right]^2 \int_{\bf q} P_0^2(q) + \lim_{k\rightarrow 0} P_{\rm line}^{1h}(k,z)  \notag \\
&- 2 \bar T_{\rm line}(z)b_1^{\rm line}(z) \lim_{k\rightarrow 0} P_{\rm line,\delta}^{1h}(k,z) + \left[\bar T_{\rm line}(z) b_1^{\rm line}(z)\right]^2 \lim_{k\rightarrow 0} P_{\rm \delta \delta}^{1h}(k,z),
\end{align}
where the low-k limit of the 1-halo terms are given by
\begin{align}\label{eq:1h_ls}
&\lim_{k\rightarrow0}  P_{\rm line}^{1h}(k,z) \hspace{.1in} = \left(\frac{c^4 p_{2,\sigma}}{4k_B^2 \nu_{\rm obs}^4 } \right)\int_{M_{\rm min}}^{M_{\rm max}} dM \ n(M,z) {\left[\frac{L(M,z)}{4 \pi \mathcal D_L^2} 
\left ( \frac{dl}{d\theta} \right )^{2} \frac{dl}{d\nu} \right ]}^2, \notag \\
&\lim_{k\rightarrow0} P_{\rm line,\delta}^{1h}(k,z) =  \frac{1}{\bar \rho_m} \left(\frac{c^2 p_{1,\sigma}}{2k_B \nu_{\rm obs}^2} \right)\int_{M_{\rm min}}^{M_{\rm max}} dM \ n(M,z)  M  \left[\frac{L(M,z)}{4 \pi \mathcal D_L^2} 
\left ( \frac{dl}{d\theta} \right )^{2} \frac{dl}{d\nu} \right ], \notag \\
&\lim_{k\rightarrow0} P_{\delta \delta}^{1h}(k,z) \hspace{.1in} =  \frac{1}{\bar \rho_m^2}\int dM  \ n(M,z) M^2,
\end{align}
with $\bar \rho$ being the average matter density in the Universe. Let us make a technical comment about evaluation of the mass integrations in Eq.  \eqref{eq:1h_ls}. While in the first two lines, the integrals have finite limits corresponding to the minimum and maximum halo masses hosting a line-emitting galaxy, the last term receives contributions from halos of all masses. In computing such integrals, however, we assume lower and upper bounds. Since the halo mass function rapidly drops at high masses, the results are rather insensitive to the upper bound. On the low mass limit we need to account for the leading-order correction due to removal of the lowest halo masses in the integral \cite{Figueroa:2012ws,Schmidt:2015gwz}. Considering the lower mass cutoff of $M_s$, and using mass conservation constraint, the last line in Eq.  \eqref{eq:1h_ls} can be approximated as
\begin{align}
P_{\delta \delta}^{1h}(k,z) &= \frac{1}{\bar \rho_m^2}\int_{M_s} dM  \ n(M,z) u(k|M)^2 M^2 \notag \\
&+ \frac{u(k|M_s)^2 M_s}{\bar \rho_m}\left[1-\frac{1}{\bar \rho_m}\int_{M_s} dM \ n(M,z) M\right],
\end{align}
which is accurate up to corrections of the order $(k R_s)^2$ with $R_s$ being the scale corresponding to the minimum mass cutoff \cite{Schmidt:2015gwz}. In our numerical evaluations, we set $M_s = 10^6 M_\odot$. 

The first term in Eq.  \eqref{eq:ls_shot_power} is the white noise due to nonlinear clustering of halos as described above, the second is the Poisson noise, and the last two terms account for halo exclusion. Depending on the relative sizes of the corrections, which depend on redshift, the shot noise may be sub- or super-Poissonian. At low redshifts of $z \leq 1$, the sub-Poissonian noise have been reported for massive halos and clusters in N-body simulations \cite{Baldauf:2013hka,Ginzburg:2017mgf,Schmittfull:2018yuk}. The super-Poisson values, on the other hand, have been observed in low-mass sub-halos of simulations. These observations align with the expectation that for massive halos, halo exclusion, which reduces the shot noise, is the leading effect. This is because the exclusion volume increases with halo mass. For low-mass halos, on the other hand, the $b_2^2$ contribution is dominant. For galaxies, the corrections also depends on the fraction of satellite galaxies; for samples with low satellite fraction that trace the centers of massive halos, sub-Poissonian noise is observed \cite{Baldauf:2013hka}. 

For intensity mapping signal, which is cumulative over all halo masses, there are various factors to keep in mind to understand the shot noise corrections. First, the dependence of the line luminosity on halo mass, which determines the relative contribution of low- and high-mass halos to line signal can change the significance of exclusion effect. Second, at redshifts $z\geq 2$, the quadratic bias $b_2$ grows rapidly with redshift (see. Fig. \ref{fig:th_bias}). Therefore, its contribution to shot-noise becomes more significant. Third, the scatter in $L(M)$ relation, enhances the Poisson shot noise more than the correction terms; the former, which is proportional to the second moment of the line luminosity is enhanced by $p_{2,\sigma}$, while the latter that is proportional to square of the first moment scales as $\left(p_{1,\sigma}\right)^2$. Lastly, the importance of exclusion terms may differ for different spectral lines, depending on the type of galaxies emitting the majority of the line signal and their satellite fraction. Since we do not implement the split of galaxies into centrals/satellites, we only investigate the first three points when comparing the prediction of the model in Eq.  \eqref{eq:shot_power} to the measured stochasticity on large scales.

\begin{figure}
    \centering
    \includegraphics[width=0.55\textwidth]{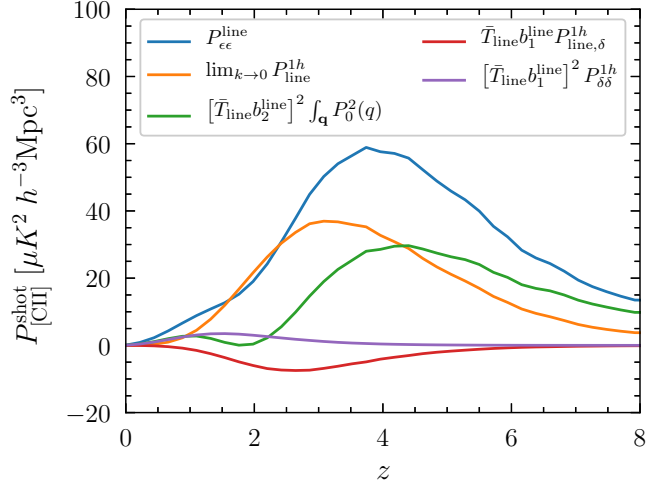}\vspace{-.2in}
    \caption{Theoretical predictions for contributions to [CII] shot noise in Eq. \eqref{eq:shot_power} as a function of redshift.}
    \label{fig:pk_stoch}
\end{figure}
To illustrate the relative sizes of the contributions to shot-noise and their redshift dependencies, in Figure \ref{fig:pk_stoch}, we show the total shot power spectrum of [CII] in blue, the Poisson noise in orange and the three corrections in green, red and purple. Since the trends for [CII] and CO shot noises are similar, we only show the results for [CII]. Several features are noteworthy in this plot; First, as also discussed in Section \ref{sec:clust}, due to peak of the star formation history at $z \simeq 2$, the Poisson shot first increase and then decrease with increasing redshift. The total shot noise, shown in blue, similarly shows a rise and fall, although the peak is shifted to higher redshifts ($z\simeq4$). Apart from a short redshift interval around $z\simeq 2$, the shot noise is super-Poissonian. For the correction term proportional to $b_2$, shown in green, the initial rise and fall at $z<2$ is due to zero crossing of the quadratic bias shown in the right panel of Fig. \ref{fig:th_bias}. At higher redshifts, the amplitude of this term grows until $z\simeq 4$, after which the drop in the amplitude of the matter power spectrum can not any more be compensated by large value of $b_2$. Finally, over most redshifts, the contribution proportional to 1-halo term of matter power spectrum (the purple line) is subdominant to the one proportional to cross line-matter 1-halo term (in red).

\section{Comparison with Simulations}\label{sec:sim_th}

In this section, we summarize the characteristics of \textsf{Mithra LIMSims} and present comparisons of the theoretical predictions with the simulated line intensity maps. Given that the implemented painting scheme does not distinguish between central and satellite galaxies and the dependence of the signal on the halo profile, our comparison only includes the 2-halo contributions for the full power spectrum shape.

\subsection{Simulation Specifications}\label{sec:sims}

The mock LIM maps presented in this work are constructed from the publicly available halo catalogs of the Hidden Valley simulation suites \cite{Modi:2019ewx} \href{http://cyril.astro.berkeley.edu/HiddenValley} {\faGlobe} \footnote{\url {http://cyril.astro.berkeley.edu/HiddenValley}}. The HV simulations have been produced with FastPM code \cite{Feng:2016yqz} \href{https://github.com/fastpm/fastpm} {\faGithub} \footnote{\url{https://github.com/fastpm/fastpm}}, and follow the dynamics of $10240^3$ particles in periodic cubic boxes of size $L_{\rm box} = 1024 \ h^{-1}{\rm Mpc}$. This gives a mass resolution of $M_{\rm min} = 8.57 \times 10^7 \ h^{-1}M_\odot$, which is sufficient to resolve halos hosting CO and [CII] up to high redshift. The accuracy of FastPM has been tested extensively in Refs. \cite{Modi:2019ewx,Feng:2016yqz}, where it was found to be accurate at the 2\% level in reproducing the clustering of halos in a full TreePM simulation.

The initial particle displacement are generated using second-order Lagrangian perturbation theory at $z_i = 99$. Halos are identified using Friend-of-Friend algorithm \cite{1985ApJ...292..371D}, implemented in FastPM code, and the halo catalogues are saved at redshifts in the range of $0.5 \leq z \leq 6$, in steps of $\Delta z = 0.5$.  The cosmology is set to a flat $\Lambda$CDM model with $\Omega_m = 0.309167, \ h=0.677, \ \sigma_8 = 0.8222, \Omega_b = 0.04903, \ n_s = 0.96834$. We perform our measurements on the \textsc{HV10240/R} simulation, which uses Gaussian initial conditions. We refer the reader to Ref. \cite{Modi:2019ewx} for more details about the HV simulations.

\begin{figure}
    \centering
    \includegraphics[width=0.495 \textwidth]{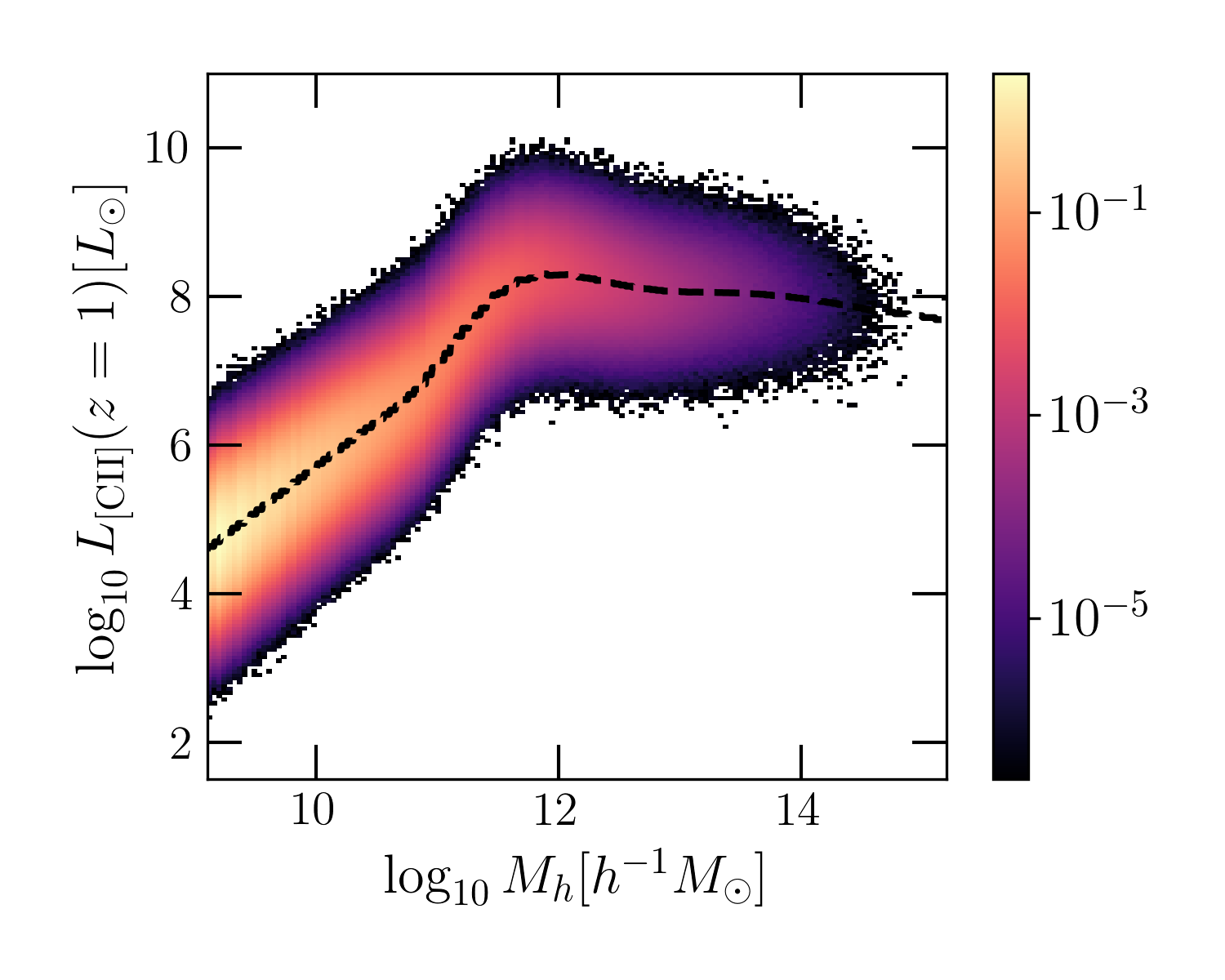}
    \includegraphics[width=0.495 \textwidth]{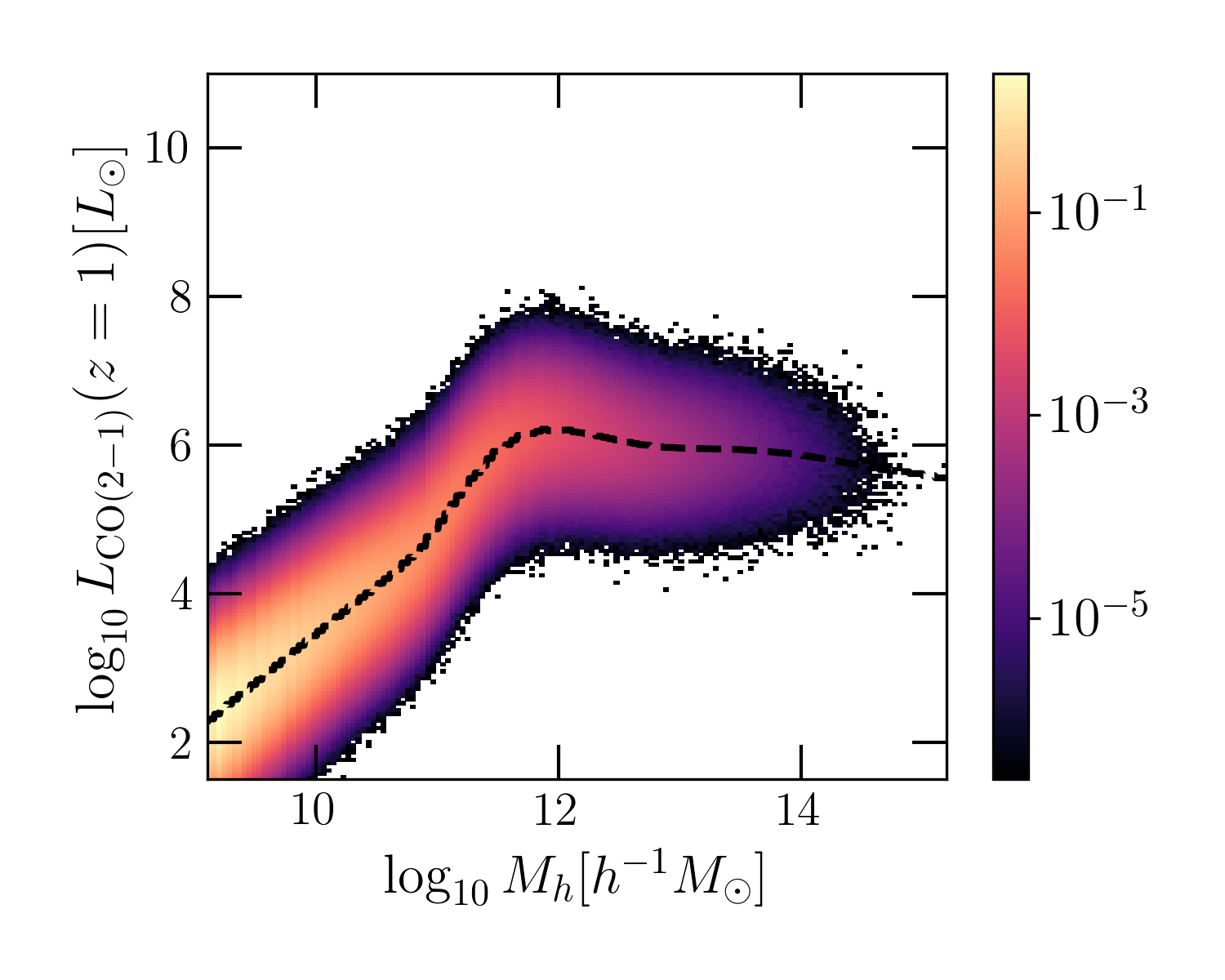}\vspace{-.1in}
    \caption{Models of [CII] (left) and CO(2-1) (right) luminosities, and the corresponding log-normal scatters with $\sigma_{\rm line}=0.37$. The colors of the scatter indicate the joint probability density of the halo mass and line luminosity.}
    \label{fig:lum_scatter}
\end{figure}
To populate the halos with CO and [CII], we extend the publicly available analysis code used in Ref. \cite{Modi:2019ewx} \href{https://github.com/modichirag/HiddenValleySims} {\faGithub} \footnote{\url{https://github.com/modichirag/HiddenValleySims}} to include the models of line luminosity as a function of halo mass and redshift, described in Eqs. (\ref{eq:SFR_IR}, \ref{eq:CO_lum_IR}, \ref{eq:CII_lum}). The measurements of the matter and line auto and cross spectra are performed using nbodykit package \cite{Hand:2017pqn} \href{https://github.com/bccp/nbodykit} {\faGithub} \footnote{\url{https://github.com/bccp/nbodykit}}. In painting halos with line luminosities, we include a halo-to-halo log-normal scatter in the relation of halo mass and line luminosity. As we discussed earlier, we consider a single scatter parameter in the relation between line luminosities and star formation rate. In Fig. \ref{fig:lum_scatter}, we show the models of line luminosities and the corresponding log-normal scatter for all halos in our simulated box at redshift $z=1$. The values of the colorbar indicate the joint probability density of the halo mass and line luminosity.

\begin{figure}[t] \vspace{-.1in}
    \centering
    \includegraphics[width=0.77\textwidth]{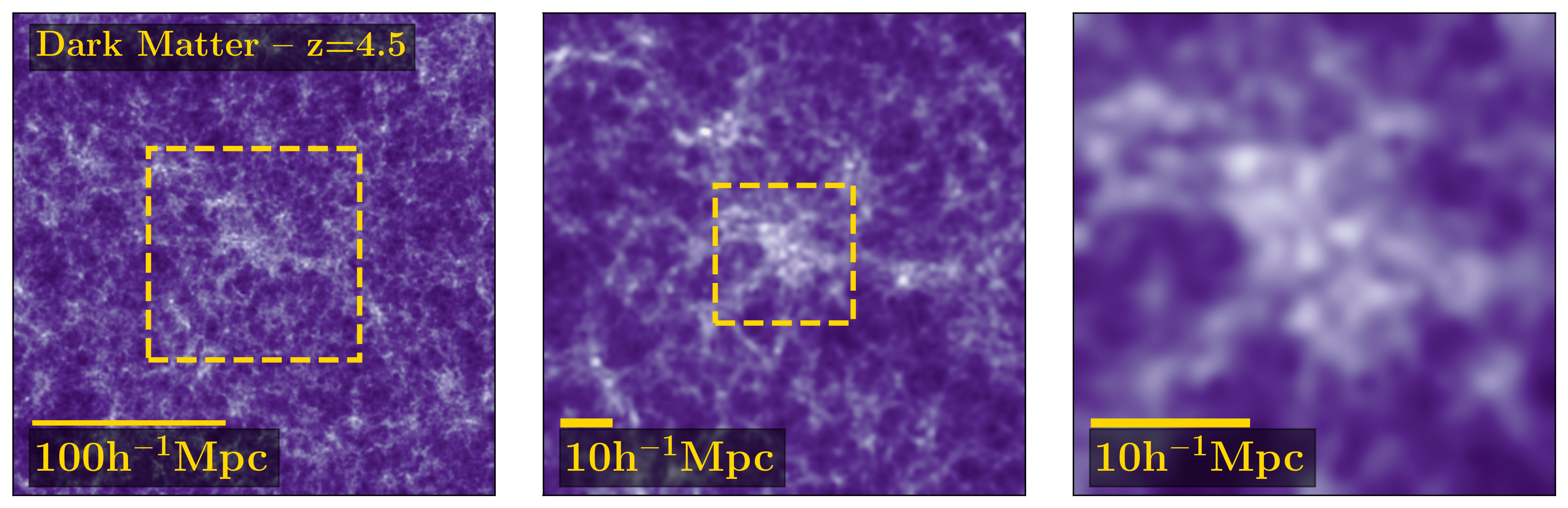}
    \includegraphics[width=0.77\textwidth]{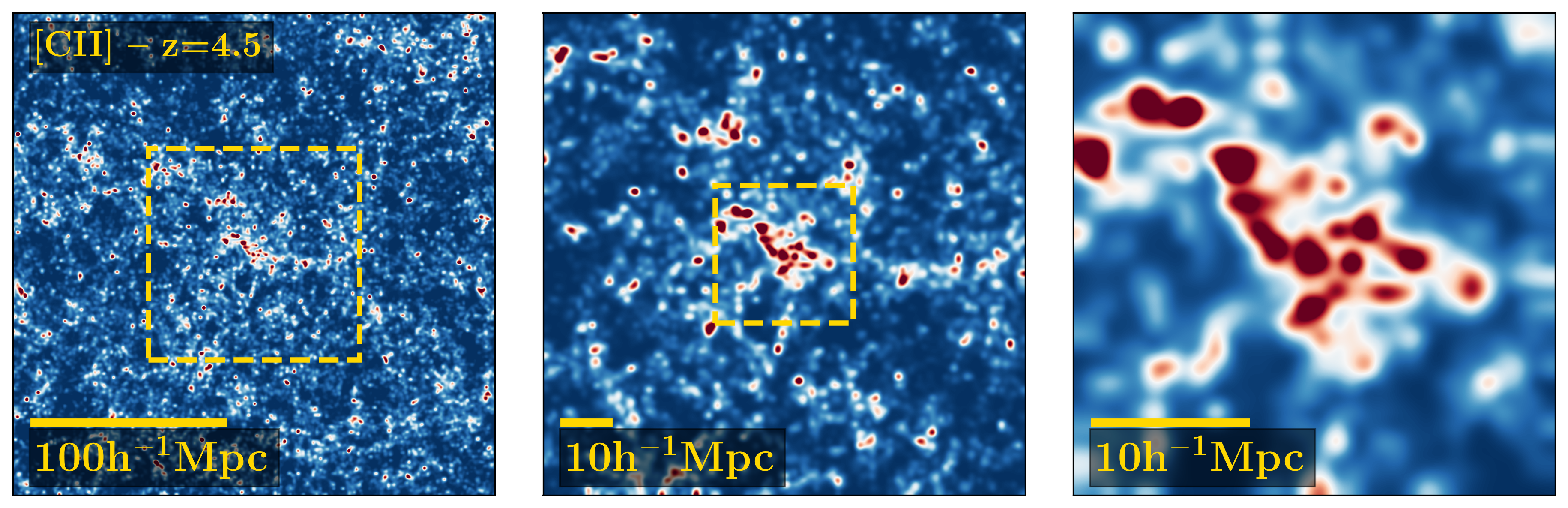} \vspace{-.07in}
    \caption{Projected cold dark matter density (top) and [CII] brightness temperature (bottom) fields at $z=4.5$. The left panels are slices of $256 \ h^{-1}{\rm Mpc}$ with the thicknesses of 64 $h^{-1}{\rm Mpc}$, centered at the position of the maximum meshed dark matter density in the full simulation box. The panels in the middle and on the right show the zoom-in slices of $128 \ h^{-1}{\rm Mpc}$ and $32 \ h^{-1}{\rm Mpc}$, and thicknesses of $32 \ h^{-1}{\rm Mpc}$ and 8 $h^{-1}{\rm Mpc}$, respectively.}
    \label{fig:mesh_z4p5}

\vspace{.25in}

    \includegraphics[width=0.77\textwidth]{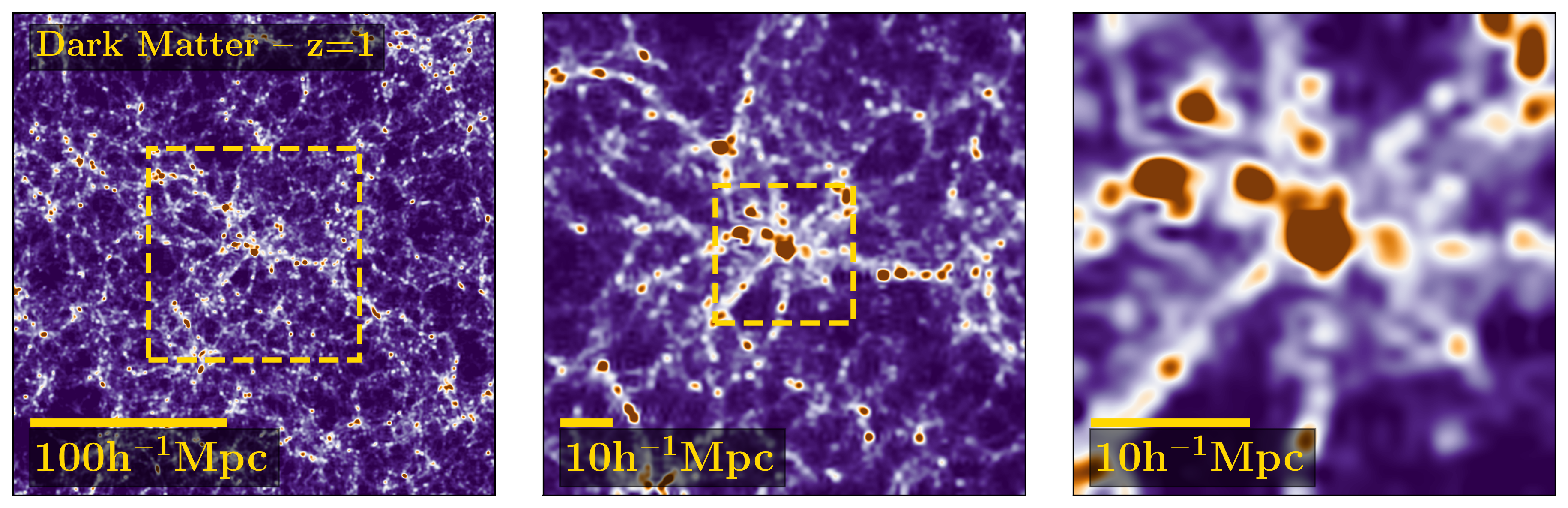}
    \includegraphics[width=0.77\textwidth]{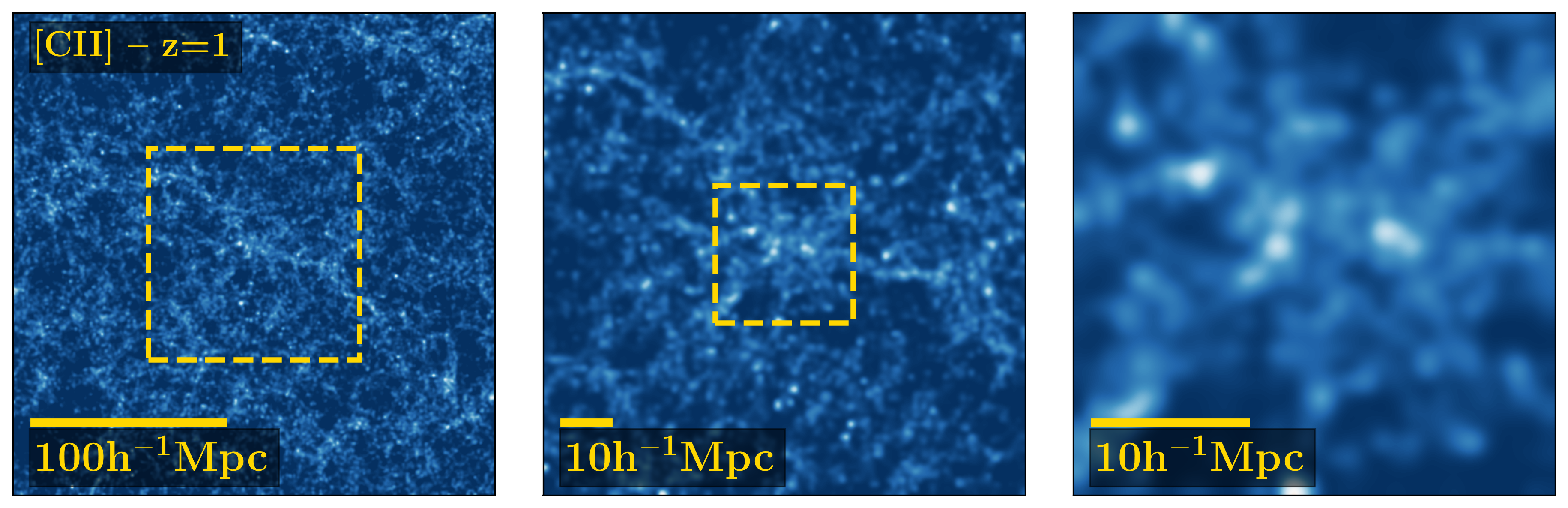}
    \includegraphics[width=0.77\textwidth]{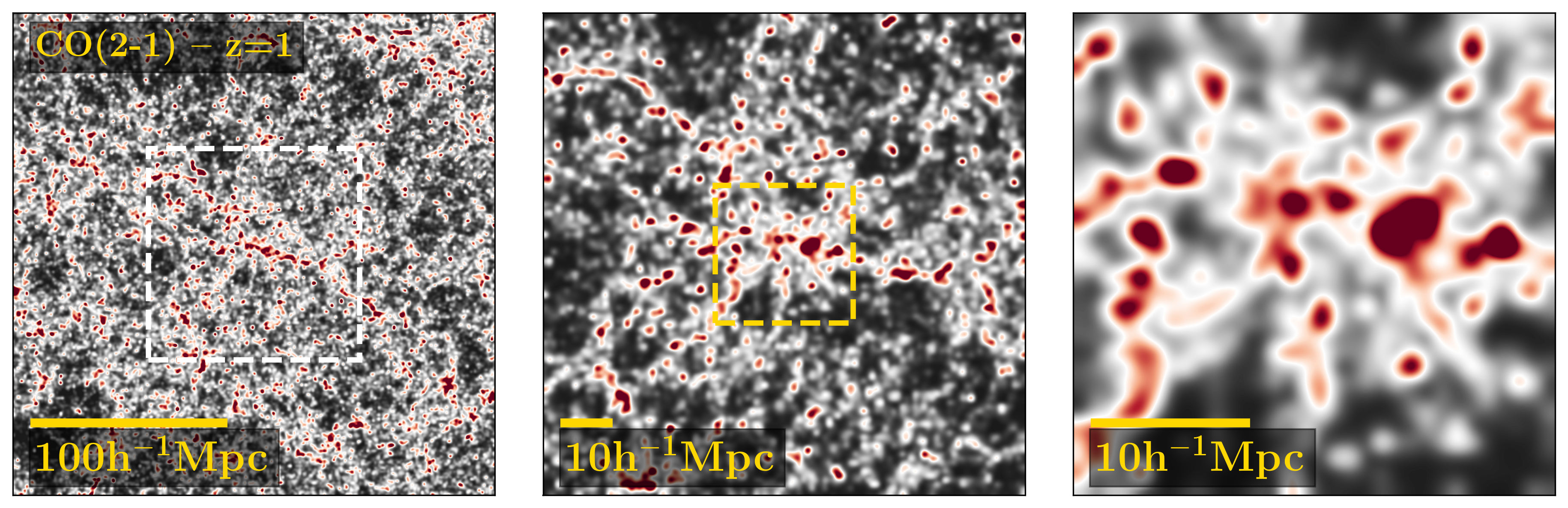}
    \vspace{-.07in}
    \caption{Same as Fig. \ref{fig:mesh_z4p5}, but now for projected cold dark matter density (top), [CII] brightness temperature (middle), and CO(2-1) brightness temperature (bottom) fields at $z=1$.}\vspace{-.2in}
     \label{fig:mesh_z1}
\end{figure}
For illustration, in Fig. \ref{fig:mesh_z4p5}, we show the projected real-space fluctuations of dark matter density (top) and [CII] brightness temperature (bottom) at redshift $z=4.5$, while in Fig. \ref{fig:mesh_z1}, we show dark-matter (top), [CII] (middle) and CO(2-1) (bottom) fluctuations at redshift $z=1$. To highlight the wide dynamic range of the simulations in mass and force resolution, the panels from left to right show slices of 256, 128, and 32 $h^{-1}{\rm Mpc}$ with the thicknesses of 64, 32, and 8 $h^{-1}{\rm Mpc}$. The slices are centered on the position of highest meshed matter density in the full simulation box of $V \simeq 1 \ (h^{-1}{\rm Gpc})^3$. The matter and [CII] fields have the same color threshold across the two redshifts. Moreover, the [CII] and CO(2-0) fields at $z=1$ have the same color threshold to depict their relative contrast. At both redshifts, the lines clearly are tracing the large-scale distribution of dark matter. At $z=1$, since [CII] has lower mean brightness temperature than CO(2-1), its projected field has clearly lower contrast. Going from $z=1$ to $z=4.5$, there is clearly less structure in the matter and line fields. On the other hand, we clearly see the rise of the temperature of the [CII] field going from $z=1$ to $z=4.5$, which is determined by the larger star formation rate at $z=4.5$ compared to $z=1$ (see e.g., Figs. 1-3 of Ref. \cite{MoradinezhadDizgah:2021upg}, showing the redshift-dependence of CO and [CII] mean brightness temperatures and power spectra). Worthy to notice in Figs. \ref{fig:mesh_z1} and \ref{fig:mesh_z4p5}, are the small-scale difference of high contrast regions in matter and line temperature fields, most clearly illustrated in the zoom-in panels on the right. The hottest spots in the line intensity fields, are associated with regions with the large number of low mass halos, which significantly contribute the observed luminosity and brightness temperature of the lines. Due to the drop and near plateauing of the line luminosities at halo masses above $M_h \simeq  10^{12} \ M_\odot/h$ (see Fig. \ref{fig:lum_scatter}), the most massive halos, which create the highest contrast regions in dark matter field, do not host a significant line emission.

\subsection{Halo Mass Function and Linear Line Bias}

\begin{figure}[t]
    \centering
    \includegraphics[width=0.55\textwidth]{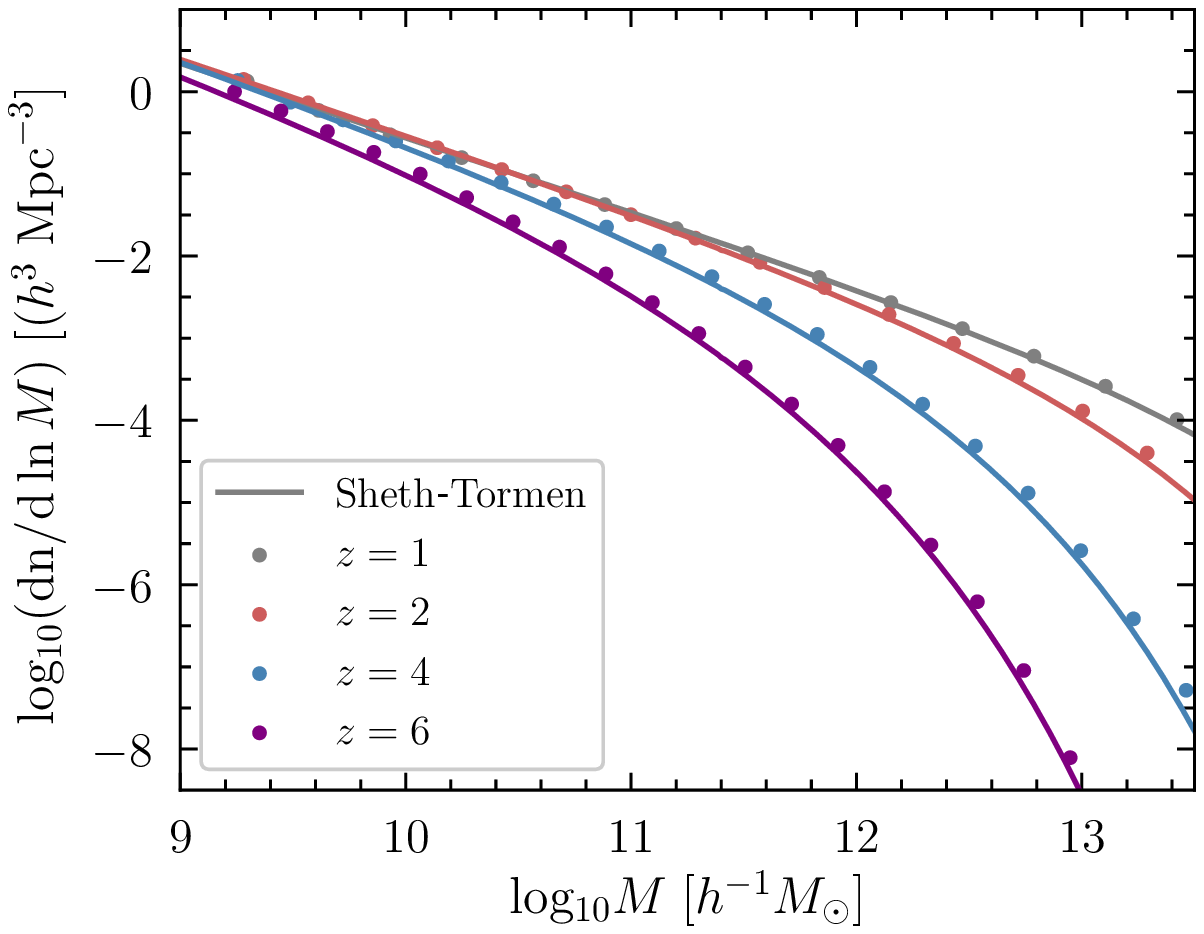}\vspace{-.15in}
    \caption{Measured halo mass function (points) and Sheth-Tormen predictions (lines), at redshifts z = \{1, 2, 4, 6\}.}\vspace{.1in}
    \label{fig:HMF}
\end{figure}

In Fig. \ref{fig:HMF}, we show the measured mass function on HV halo catalogs (points) vs. the ST mass function (lines) at redshifts $ z = \{1, 2, 4, 6\}$. The theoretical predictions lie below the measured mass functions for all halo masses for the two highest redshifts. The data points are slightly below the ST predictions for the two lower redshifts. For $z=1$, this under-prediction extends over a wider range of halo masses. As we will see later in this section, these offsets, in turn, would result in a discrepancy of $\leq10\%$ in predicted and measured linear bias of the lines. Similarly, the offset in the mass functions results in differences in mean brightness temperature and Poisson shot noise. When fitting the full shape of the power spectrum in HV simulations, we compensate for the offsets by re-scaling the linear bias of the line and using the measured Poisson shot noise instead of the theoretical prediction. 

\begin{figure}[t]
    \centering
    \includegraphics[width=0.49\textwidth]{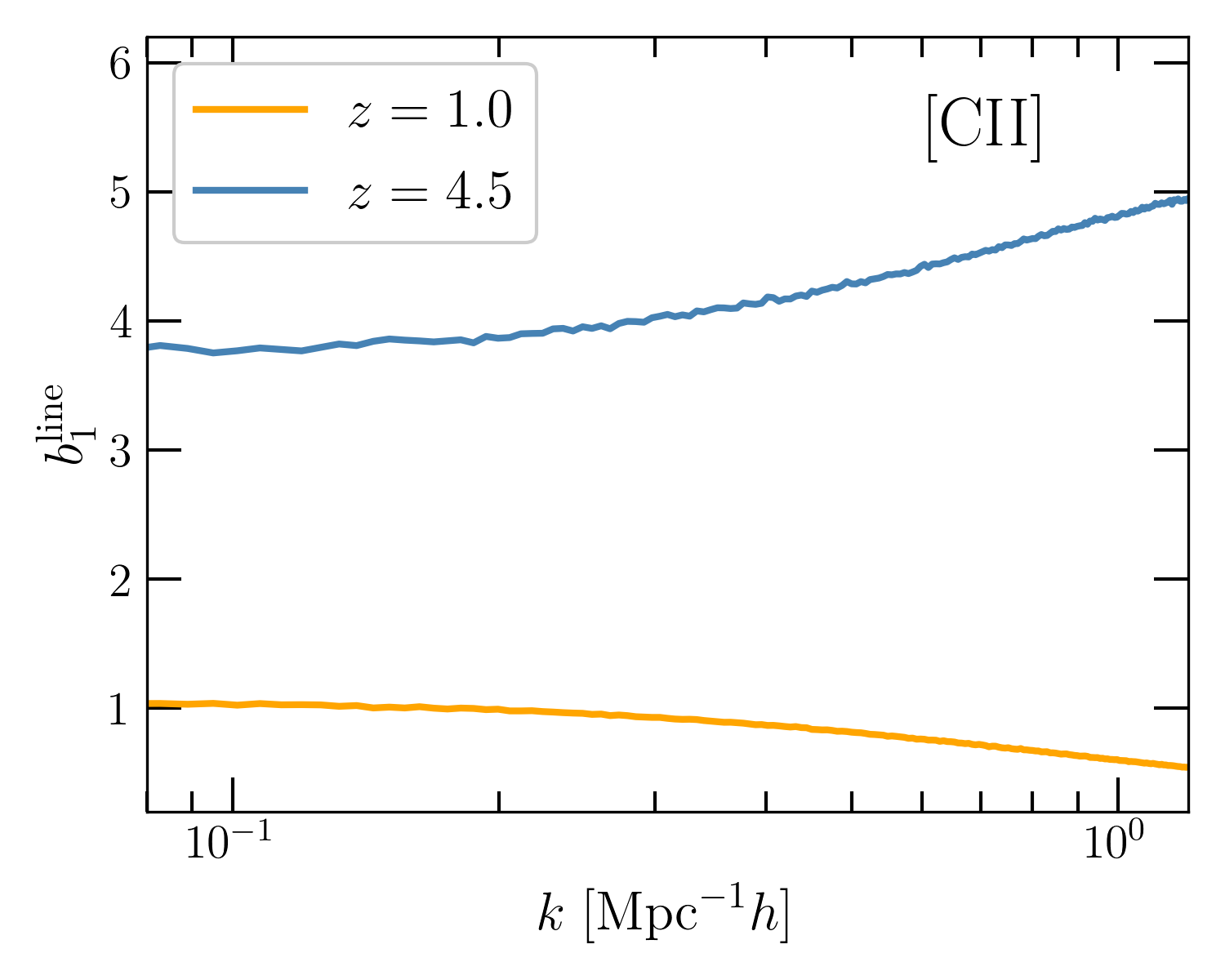}   
    \includegraphics[width=0.49\textwidth]{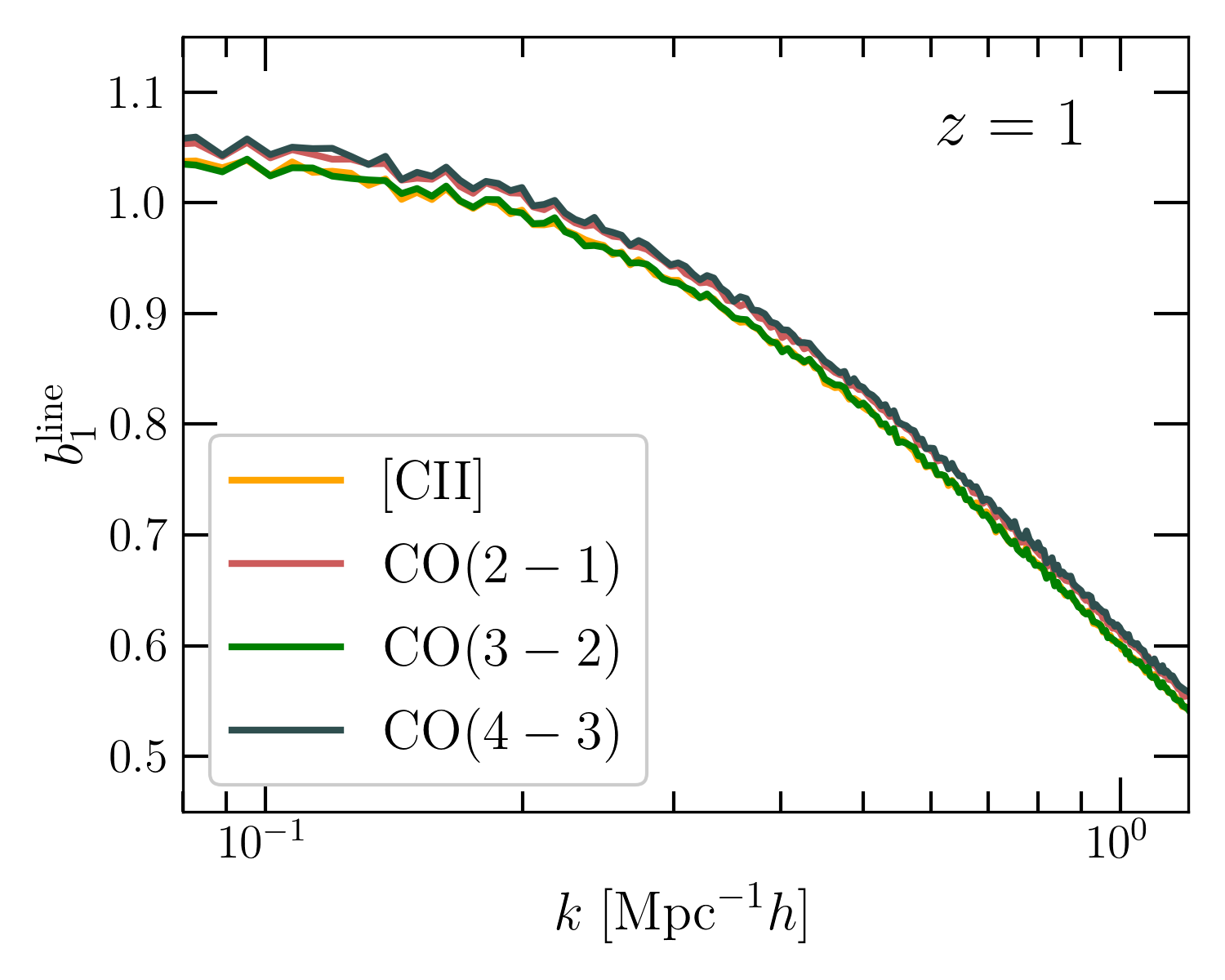}\vspace{-.1in}
    \caption{{\it Left panel:} Linear bias of [CII] emission lines at redshifts $z=1$ (in orange) and $z=4.5$ (in blue). {\it Right panel:} Linear bias of three CO rotational lines (J=2-1 to J=4-3) at $z=1$.}
    \label{fig:sim_bias}
\end{figure}

In Fig. \ref{fig:sim_bias}, we show the linear bias of [CII] at $z = \{1, \ 4.5\}$ on the left, and three of the CO lines at $z=1$ on the right. The linear biases are measured from the ratio of the matter-line cross-spectrum to matter power spectrum (hat refers to measured quantities)
\be
\hat b_1^{\rm line} = \frac{1}{\hat{\bar{T}}_{\rm line}} \left[\frac{\hat P_{{\rm line},\delta}(k)}{\hat P_{\delta \delta}(k)} \right].
\ee
We measure the mean brightness temperature of the line by evaluating the first moment of line luminosity, averaged all halo masses in HV catalogue with masses in the range of $[M_{\rm min},M_{\rm max}]$, 
\be
\hat {\bar T}_{\rm line} = \left( \frac{c^3}{8 \pi k_B \nu_{\rm obs}^3 H(z)} \right)\frac{1}{L_{\rm box}^3} \sum_h {\hat L}(M_h,z).
\ee 
Similarly, we measure the Poisson shot noise of the lines by computing the normalized mass-averaged second moment of the line luminosity, 
\be
{\hat P}_{\rm shot}(k,z) = \frac{ L_{\rm box}^3 \sum_h {\hat L}^2(M_h,z)}{\left[\sum_h {\hat L}(M_h,z)\right]^2}.
\ee
As we go to smaller scales, the biases become clearly scale dependence, indicating the necessity of introducing higher-order line biases. At $z=1$ the scale-dependence of CO and [CII] linear biases show the same trend (decreasing with increasing $k$). Since the quadratic line bias grows rapidly with increasing redshift at $z>2$ (see the right panel of Fig. \ref{fig:th_bias}), the linear bias of [CII] at $z=4.5$, instead, increases with increasing $k$.   

\subsection{Full Shape of Line Power Spectrum}

\begin{figure}[t] \vspace{-.1in}
    \centering
    \includegraphics[width=0.49\textwidth]{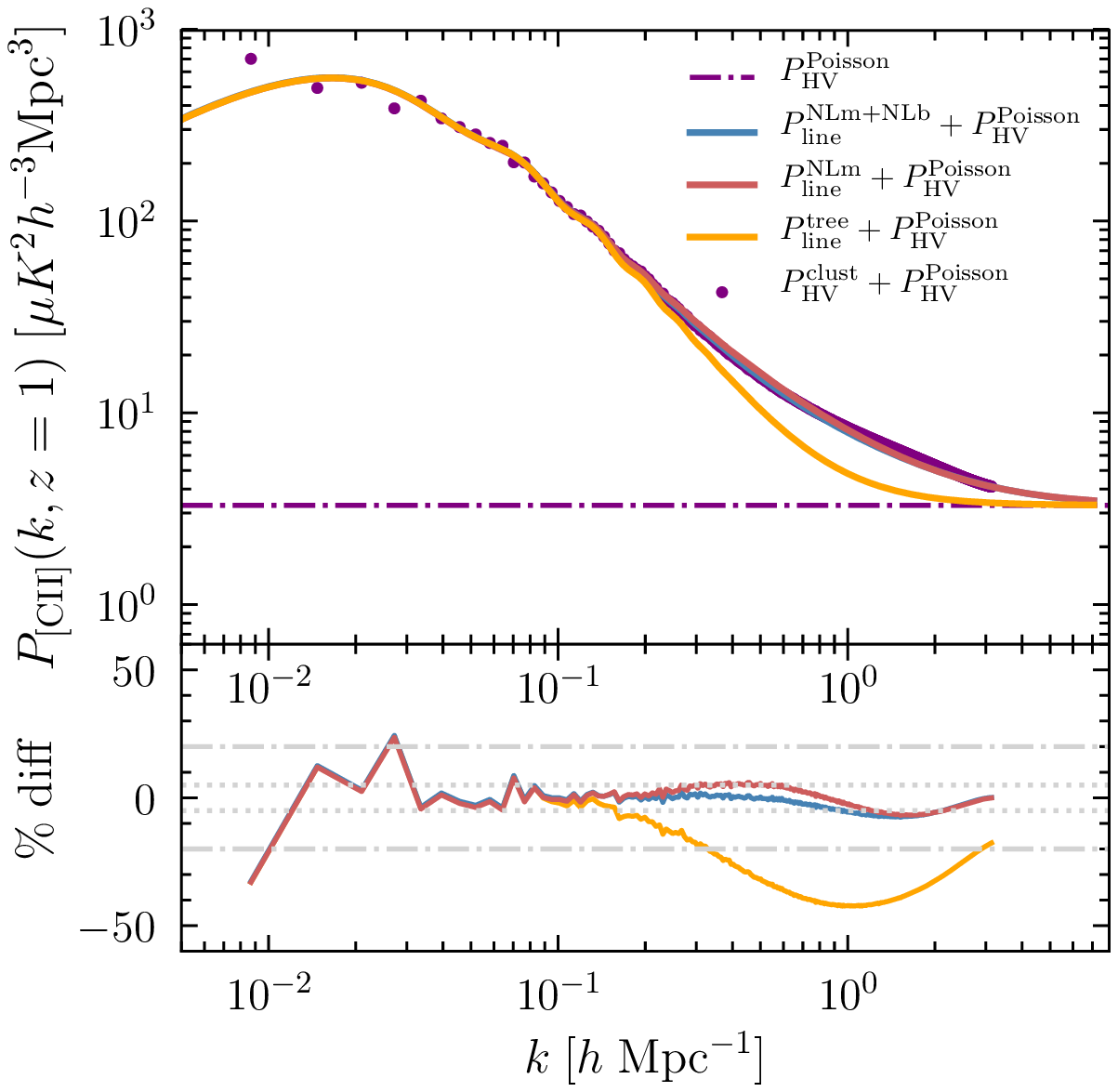}
    \includegraphics[width=0.49\textwidth]{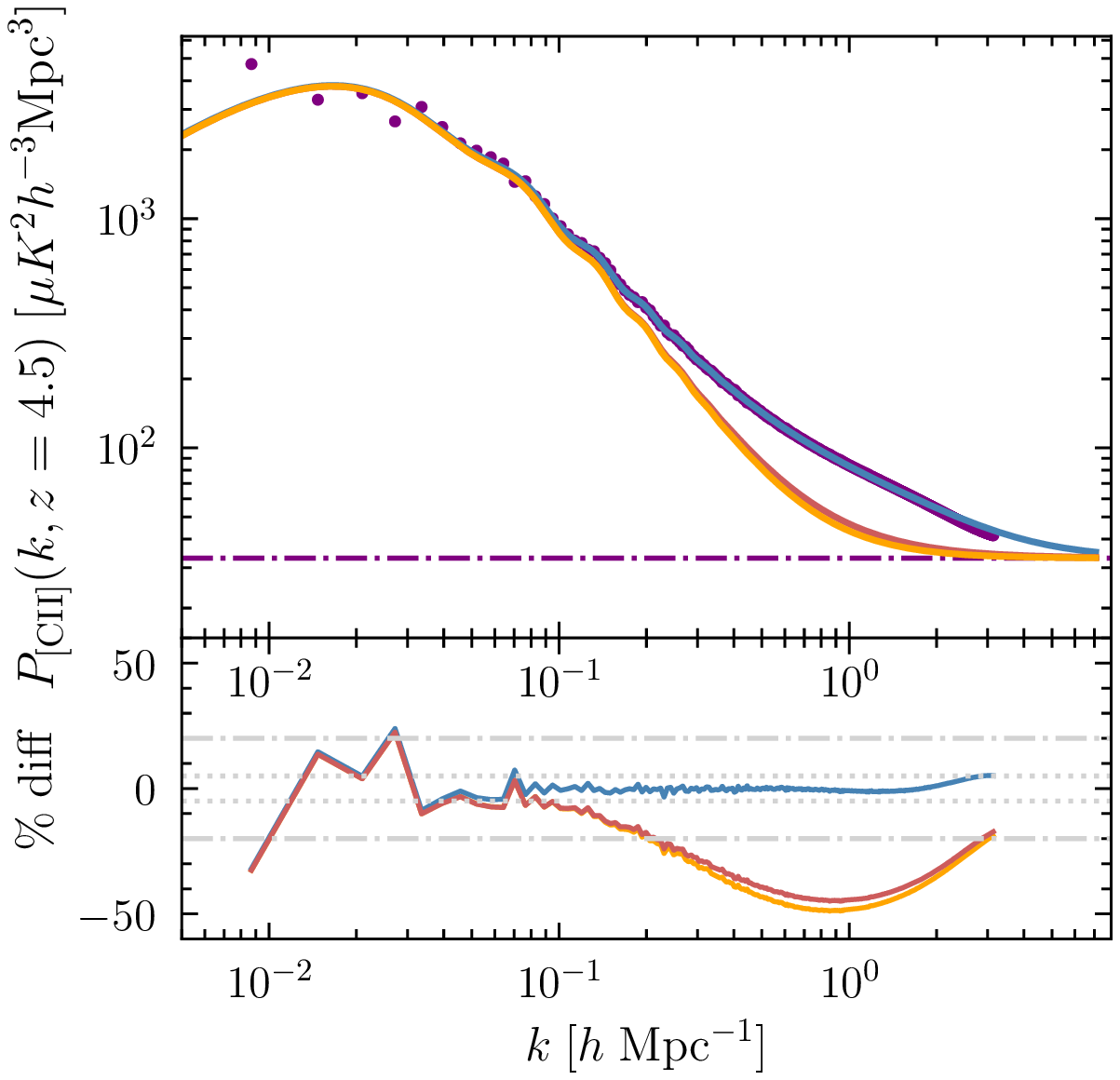}\vspace{-.1in}
    \caption{{\it Top panels:} Power spectra of [CII] power spectrum at $z=1$ on the left and $z=4.5$ on the right. The horizontal dashed-dotted line is the measured Poisson shot noise. The points are the measured line power spectrum, while the solid lines are the model variations of the 2-halo term that we consider; in yellow the tree-level, in red the one-loop in matter, and in blue the one-loop in matter and bias.  In all cases, we have added the measured Poisson noise to the clustering power, $P_{\rm HV}^{\rm Poisson}$. {\it Bottom panels:} Fractional difference between the measured and theoretical power spectra. The horizontal dotted and dashed-dotted lines in the bottom panel of each plot indicate the 5\% and 20\% fractional differences.}
    \label{fig:sim_pk_hm}
\end{figure}

In Fig. \ref{fig:sim_pk_hm}, we show the measurements of the total (clustering + shot) [CII] power spectrum (purple points) vs. the theoretical predictions (solid lines). The plot on the left corresponds to $z=1$, and the one on the right to $z=4.5$. In each plot, the bottom panel shows the fractional difference between the measurement and the theoretical model. We show the three variations of the clustering contribution of line power spectrum; linear matter and bias, $P_{\rm line}^{\rm tree}$ (in yellow), linear bias and one-loop matter, $P_{\rm line}^{\rm NLm}$ (in red), and one-loop for both matter and bias, $P_{\rm line}^{\rm NLm + NLb}$ (in blue). The IR resummation is included in both the red and blue curves, while the $\tilde b_{\nabla^2}$ is included only in the blue curve. The purple dashed-dotted line corresponds to the measured shot-noise. The values of quadratic and cubic biases are fixed to the theoretical values, the theoretical prediction of the linear bias is rescaled to match the measured ${\hat b}_1^{\rm line}$, and the value of the effective parameter, $\tilde b_{\nabla^2}$, is fitted by eye for each line at a given redshift. In all cases, the measured Poisson shot noise is used in the fits. Table \ref{tab:vals}, shows the values of these three values for CO and [CII] lines. For [CII], all three values increase as we go to higher redshifts. At $z=1$ the values of the bias parameters are nearly the same for all of the lines. 

Interestingly, even with this 1-parameter fit (together with re-scaling of the bias), we can achieve an agreement between the measured and predicted power spectra at better than $5\%$ on scales $k \lesssim 0.8  \ h/{\rm Mpc}$ at $z=1$ and $k \lesssim 2 \ h/{\rm Mpc}$ at $z=4.5$. For CO rotational lines, we obtain similarly good fits. We can identify the following trends in Fig. \ref{fig:sim_pk_hm}; first, the tree-level model of the line power spectrum significantly under-predicts the line power spectrum at both redshifts. At $z=1$, accounting for the loop contributions of matter power spectrum, improves the fit considerably. Adding the bias loops improves the fit (on the scales of $0.2 \lesssim k \lesssim 0.8 \ [h/{\rm Mpc}] $) only slightly. This, of course, is not unexpected, since at $z=1$ the nonlinear line biases are still small, while matter fluctuations are highly nonlinear. Including the $k^2P_0(k)$ term in the one-loop matter power spectrum, would improve the fit for the red line. At $z=4.5$, on the other hand, the matter fluctuations are much more linear, while the bias loops get large as values of nonlinear biases become larger. Therefore, the model with one-loop matter does nearly as bad as the tree-level model, while the full one-loop model provides an excellent fit to the data.  

\begin{table}[ht!]
    \par\smallskip
    \centering
    \begin{tabular}{c c c c c}
    \hline \hline
    \xrowht{20pt}
    ${\rm line}$ & ${\rm redshift}$ & ${\hat b}_1^{\rm line}$ & ${\hat {\tilde b}}_{\nabla^2}$ & $P_{\rm HV}^{\rm Poisson} $ \\  \hline 
    ${\rm [CII]}$ & 1.0   & 1.04 & 0.6 & 3.29 \\
    ${\rm [CII]}$ & 4.5 & 3.78 & 2.9 & 33.0 \\
    CO(2-1) & 1.0 & 1.05 & 0.6 & 65.7\\
    CO(3-2) & 1.0 & 1.04 &  0.6 & 34.8  \\
    CO(4-3) & 1.0 & 1.06 & 0.6 & 11.6\\
    \hline
    \end{tabular}
    \caption{Values of the measured linear bias, ${\hat b}_1^{\rm line}$, the effective parameter, ${\hat {\tilde b}}_{\nabla^2}$, and the Poisson shot-noise, $P_{\rm HV}^{\rm Poisson} $, used in the fits of the line clustering power spectrum. The shot-noise is in unit of $[\mu K^2 h^{-3} {\rm Mpc}^3]$).}
    \label{tab:vals} 
\end{table} 

In the context of existing and upcoming pathfinder LIM surveys (e.g., \cite{Li:2015gqa,doi:10.1117/12.2057207,Lagache_2018}), which measure the line intensity power spectrum in the clustering regime, neglecting the nonlinear contributions to matter fluctuations and biasing results in $\geq 20\%$ discrepancy between the theoretical prediction and the measured power spectrum on scales $k \geq 0.3 \ {\rm Mpc}^{-1}h$ (at $z=1$). The validity regime of the linear model is further reduced at higher redshifts. One has to keep in mind though that in order to fully determine the validity regime of the model, several observational effects should be further included; Above all, since the line power spectrum is observed in redshift-space, accounting for the distortions due to peculiar velocities of the line emitters, i.e. redshift-space distortions is necessary. In general the linear model of the power spectrum of the biased tracers in redshift-space fails at lower $k$-values than the real-space one. In addition to RSD, other astrophysical effects, such as quenching factor of galaxies (which we assumed to be unity), the scatter in the line-luminosity-halo mass (which we assumed to be constant), and broadening of spectral lines, in particular if they depend on the halo masses, can additionally modify the shape of line power spectrum and in turn affect the significance of non-linear corrections. However, it is worth noting that such effects should affect shot-noise contribution more significantly than the clustering part (see e.g.,  \cite{COMAP:2021rny}), since they are expected to be more relevant for brightest line emitters, which are not the dominant contributors to clustering. Therefore, one does not expect that they diminish the importance of modeling of the nonlinearities in 2-halo term.

\subsection{Stochasticity on Large Scales}

\begin{figure}[t]
    \centering
    \includegraphics[width=0.49\textwidth]{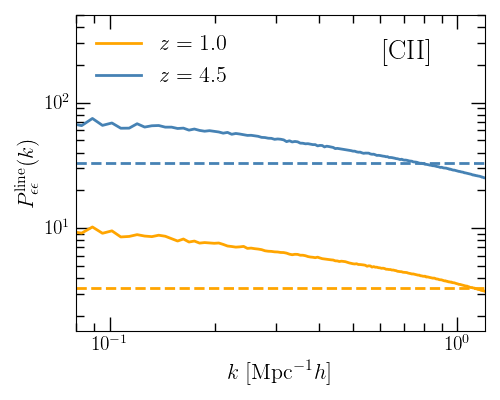}   
    \includegraphics[width=0.49\textwidth]{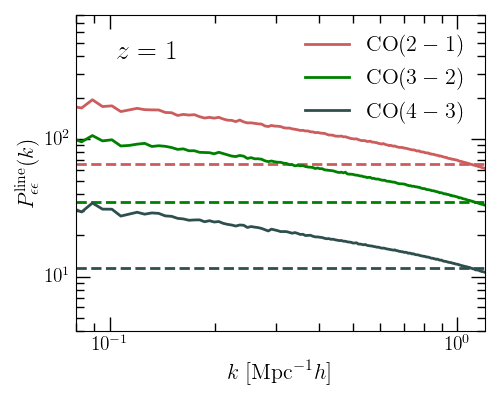} \vspace{-.1in}
    \caption{Measured stochasticity power spectrum (solid lines) versus the Poisson shot noise (dashed lines) for [CII] (left panel) at redshifts $z=1$ (red) and $z=4.5$ (in green) and for the three CO rotational lines (with J=2-1 to J=4-3) at $z=1$.}
    \label{fig:sim_stoch}
\end{figure}

Next, we study the significance of non-Poisson corrections to shot noise on large scales. Based on the definition of stochastic fields on large-scales in Eq.  \eqref{eq:stoch}, we measure the shot power using measurements of line and matter power spectra and their cross-spectrum,
\be\label{eq:pk_stoch}
P_{\epsilon \epsilon}^{\rm line}(k) = \hat P_{\rm line}(k) - 2 {\hat {\bar T}}_{\rm line}\hat b^{\rm line}_1 \hat P_{{\rm line},\delta}(k) + {\hat {\bar T}}_{\rm line}^2 (\hat b^{\rm line}_1)^2 \hat P_{\delta \delta}(k)
\ee
In Fig. \ref{fig:sim_stoch}, solid lines show the measured shot power for [CII] line at $z=1$ (on the left) and at $z=4.5$ (on the right). The dashed lines are the Poisson shot noise. In agreement with theoretical predictions (shown in Section \ref{sec:th_shot}), for [CII] at both redshifts, the shot is super-Poissonian. The same trends are also seen for CO lines. As described previously, given the slight discrepancy between the theoretical and measured mass functions, there is a slight offset between the measured and predicted stochasticity.

Let us close this subsection by noting that in constraining cosmological parameters from the full shape of the line power spectrum, instead of computing the corrections to the Poisson shot noise, one can treat the deviation as a nuisance parameter and marginalize over it in the likelihood analysis. This is now a common practice in galaxy clustering analysis (see e.g. Ref. \cite{Gil-Marin:2020bct}). In the context of astrophysics and constraining star-formation history from measurements of the shot and clustering power, however, the deviations from Poisson noise changes the relation of the observed shot noise to line luminosity and hence to star formation history. Therefore, to obtain astrophysical constraints from LIM observation, possible non-Poissonian nature of the shot noise can not be neglected.

\begin{figure}[t]
    \centering
    \includegraphics[width=0.72\textwidth]{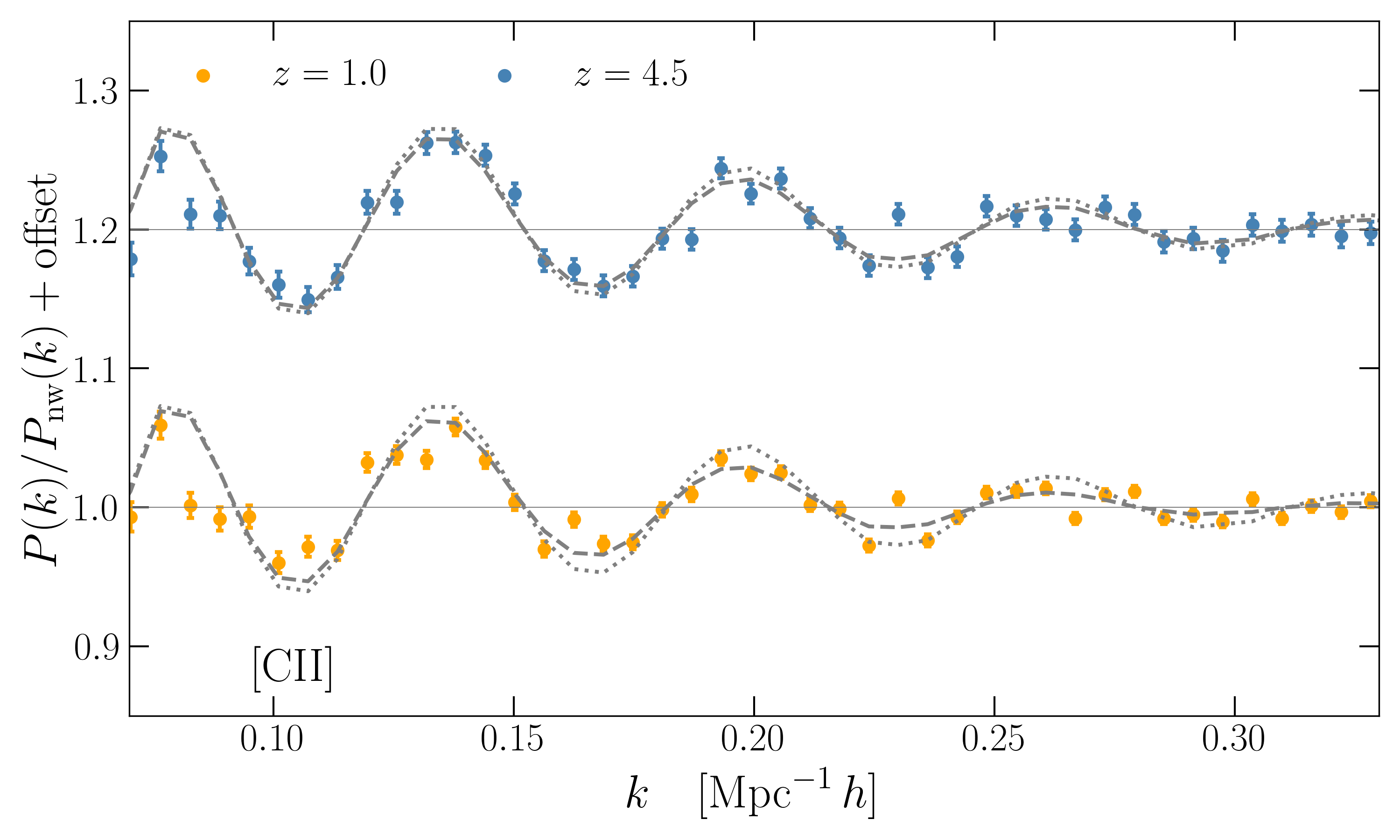}\vspace{-.1in}
    \caption{The BAO signal extracted by taking the ratio of the line power spectrum to the smooth broadband component. The dotted lines are the theoretical predictions using linear theory, while the dashed lines are the nonlinear model. The vertical lines are the expected error bars for the BAO measurement in [CII] intensity maps at $z=1$ (bottom curve) and $z=4.5$ (top curve) from space-mission as proposed in EASA Voyage 2050 Microwave Spectro-Polarimetry white paper \cite{Delabrouille:2019thj}. The curves at different redshifts are shifted w.r.t. each other to make them visually easier to distinguish.}
    \label{fig:bao_CII}

    \vspace{.5in}
    
    \includegraphics[width=0.72\textwidth]{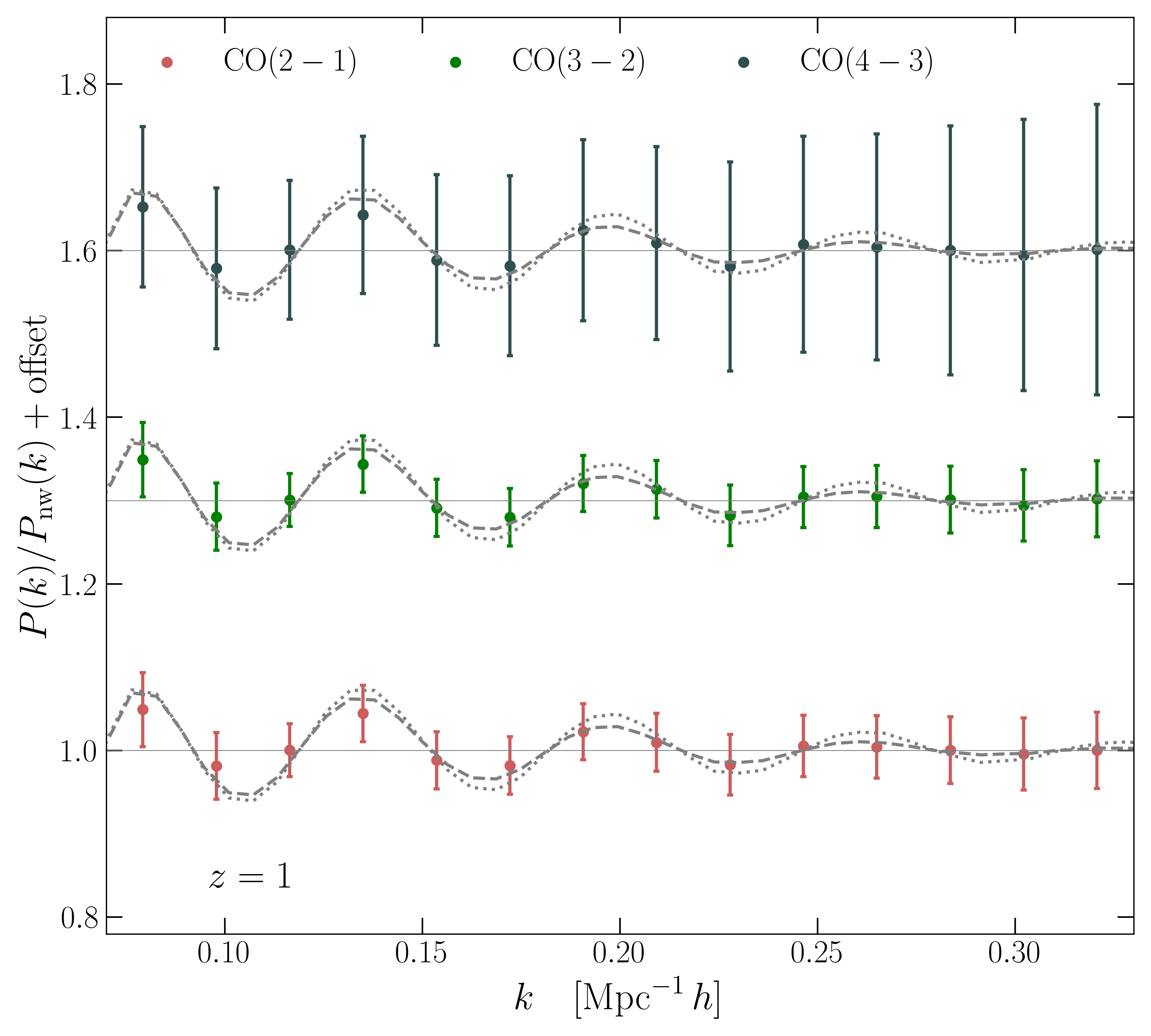}\vspace{-.1in}
    \caption{Same as Fig. \ref{fig:bao_CII}, but for CO rotational lines, with the error bars corresponding to ground-based SPT-like survey with $\sim 10^6$ spectrometer-hours. The curves from top to bottom correspond to CO(4-3) in black, CO(3-2) in green, and CO(2-1) in red. Again, the curves are shifted w.r.t. each other to make them visually easier to distinguish.}
    \label{fig:bao_CO}
\end{figure}

\subsection{Baryon Acoustic Oscillations}

To highlight the importance of accounting for the smoothing of the BAO signal due to large bulk velocities, i.e., the IR resummation, in Fig. \ref{fig:bao_CII}, we show the BAO signal extracted from [CII] power spectrum by dividing it by the broadband (smooth component) at $z=1$ and $z=4.5$. Similarly, we show the BAO signal of the three CO line at $z=1$. In both figures, the curves are shifted w.r.t each other to make them easily distinguishable. The points are the measurements on simulated maps, the dotted lines are the tree-level model, and the dashed lines are the full nonlinear model. The vertical lines are the projected observational errorbars that we discuss below. As expected, the smoothing of BAO is more significant at lower redshifts, where it is also clear that the nonlinear model provides a better fit to the BAO signal (in particular at $k\lesssim 0.2 \ h/{\rm Mpc}$). The fit can be further improved by fitting the line biases instead of fixing them to their theoretical values. 

The distance measurement using the BAO method \cite{Weinberg:2013agg} as a function of redshift, which can be uniquely performed by LIM surveys, provide an exciting prospect to shed light on the origin of cosmic acceleration and potential deviation of general relativity. To illustrate the potential of future mm-wavelength LIM surveys to measure the BAO signal, in Figs. \ref{fig:bao_CII} and \ref{fig:bao_CO}, we also show the expected error bars for future LIM surveys. The error estimates include the shot noise, instrument noise and sample variance for the volume corresponding to redshift bins with width of $\Delta z/(z+1) = 0.1$ dex around the central redshift.  As a representative of near-term future survey, capable of early detection of the BAO feature, for the CO lines at $z=1$, we consider a hypothetical ground-based mm-wavelength survey with a relatively modest number of spectrometer-hours ($\sim 10^6$), covering two fields of 100 square degrees apiece, using spectrometers with 1000 km/sec ($R=300$) channel resolution and background-limited performance in astral winter weather condition (matching that in Ref. \cite{MoradinezhadDizgah:2021upg}). It is worth noting that a multi-year survey with SPT-SLIM\footnote{The South Pole Telescope Summertime Line Intensity Mapper (SPT-SLIM) will demonstrate a novel spectrometer-on-a-chip technology, to be deployed in 2023.} can hypothetically achieve this number of spectrometer hours \cite{Karkare:2021ryi}. Detecting CO lines at higher redshift is more challenging; as $z=1$ is an optimal point for the balance of instrument sensitivity and signal strength. To overcome the low signal and higher instrument noise at higher redshifts, one needs higher spectrometer-hours. So we are only showing $z=1$ as early detection possibility.  As an example of a more futuristic survey, capable of measuring BAO at percent-level precision, given the challenges of ground-based measurements of [CII] due to limited atmospheric transmission at the relevant frequencies, we consider a space-based mission with a similar specification to the design proposed in Refs. \cite{Delabrouille:2019thj, Silva_2017}, with 64 spectrometers with spectral resolution of $R=300$ and background-limited performance and an actively cooled aperture, with a survey covering $f_{\rm sky}=0.25$ in a total of 18000 hours. We refer the reader to Refs. \cite{MoradinezhadDizgah:2021upg, Delabrouille:2019thj} for details of instrument specifications and noise calculation.

\section{Conclusions}\label{sec:summary}

Line intensity mapping offers exciting prospects for precision tests of cosmology and fundamental physics by mapping the large-scale structure over a broad range of spatial scales and redshifts. In addition to tackling observational challenges, i.e. control of systematics and accurate removal of foregrounds, realizing the potential of LIM surveys in constraining cosmology is contingent upon having accurate theoretical predictions of the clustering statistics of line fluctuations. In order to assess the accuracy of the theoretical models of the line, cosmological-scale simulations of line intensity are essential. The challenge in generating simulated intensity maps, however, is the simultaneous necessity of having large volume to study clustering on large scales and high mass resolution to capture the contributions of low-mass halos to the line signal.   
In this paper, we presented the first cosmological simulations of CO and [CII] intensity fluctuations (periodic box of $L_{\rm box} \sim 1 \ h^{-1}{\rm Gpc}$ side and halo mass resolution of $M_{\rm min} \sim 10^8 \ h^{-1} M_\odot$), which are produced using the publicly available halo catalogs from HV simulations. To populate the halos with line intensity we assume the commonly applied scaling relations and use the empirical fit for the dependence of star-formation rate, halo mass and redshift. 

Focusing on line intensity power spectrum, we investigated the importance of accounting for nonlinear effects in relating the properties of halos to the underlying dark matter distribution, and the deviations from Poisson shot noise. Using an extended halo model, which accounts for nonlinear perturbative corrections to halo power spectrum in the 2-halo term using EFTofLSS, we showed that at both $z=1$ and $z=4.5$, neglecting the nonlinearities in dark matter distributions and biasing relations of line intensity, underestimates the amplitude the signal significantly (by $>20\%$ for $k\geq 0.3 \ {\rm Mpc}^{-1}h$ at $z=1$ and for $k\geq 0.2 \ {\rm Mpc}^{-1}h$ at $z=4.5$). At lower redshift, nonlinearities of dark matter fluctuations play a more significant role, while at higher redshifts accounting for nonlinear line biases is essential. Fitting only a single parameter (parameterizing the amplitude of the $k^2P_0(k)$ contribution to the line power spectrum and setting the value of all the nonlinear biases to their theoretical predictions (and re-scaling the linear bias to match the measured value), the model reaches an agreement of better than 5\% with the measured line power spectrum. We showed that in agreement with theoretical predictions, the shot noise of the CO and [CII] lines are (almost always) super-Poissonian in the redshift range considered. At $z>2$, this deviation is largely driven by the contribution proportional to quadratic bias $b_2$, which grows large with increasing redshift. 

Lastly, we discussed the nonlinear smoothing of the BAO signal due to large bulk velocities, and showed that the one-loop EFT-based halo-model describes the BAO measurement in simulated CO and [CII] power spectra accurately. As expected, the nonlinear smoothing of the BAO is more significant at lower redshift. To highlight the prospects of the first detection and precision measurements of BAO signal by future LIM surveys probing CO and [CII], we also presented expected errorbars for near- and long-term future BAO measurement of three CO lines at $z=1$ by a ground-based survey, and of [CII] line at $z=1, 4.5$ by a space-based survey. 

This work is only the first step in constructing and testing the theoretical models of statistical descriptors of line intensity fluctuations, at the precision level needed for constraining cosmology with future LIM surveys. There are several directions in which this study can be extended; The next natural step is to include the redshift-space distortions in the modeling of halo-matter relation, and to investigate the impact of non-linearties on multipoles of line power spectra. Additionally, the HOD model of the line power spectrum should be extended to distinguish the contribution of central and satellite galaxies to line signal. This distinction can for instance alter the small-scale suppression of the redshift-space power spectrum due to Finger-of-God. Furthermore, to improve the modeling of line-halo relation, the empirical semi-analytic scaling relations, describing the line luminosity, should be replaced by a physically motivated model capable of self-consistently predicting various line emissions originating from different phases of the interstellar (ISM) medium. It is essential to have a model that incorporates relevant ISM physics, yet is simple enough that can be efficiently used in theoretical predictions of the line signal and in generating mock intensity maps based on Nbody simulations. In this regard, sophisticated simulations of hydrodynamical galaxy formation and radiative transfer (e.g., \cite{Pallottini:2019uil,Leung2020,Kannan:2021ucy}) together with observational data (e.g., cosmic infrared background anisotropies \cite{Sun2019}, and galaxy UV luminosity functions), provide guidelines for constructing such models.

\section*{Acknowledgement}

It is our pleasure to thank Vincent Desjacques for illuminating discussions on halo stochasticity, Chirag Modi for discussions related to HV simulations, and Jiamin Hou for input regarding visualization of simulated projected fields. We also thank Vincent Desjacques, Adam Lidz, and Anthony Pullen for their helpful feedback on the draft of this manuscript. A.M.D. is supported by the SNSF project ``The  Non-Gaussian  Universe and  Cosmological Symmetries", project number:200020-178787. A.M.D also acknowledges partial support from Tomalla Foundation for Gravity.

\appendix

\section{Explicit Forms of One-loop Integrals}\label{app:1loop}

The matter power spectrum up to one-loop in standard perturbation theory is given by \cite{Bernardeau:2001qr} 
\bea
P_m^{\rm 1-loop}(k)  = P_m^{(22)}(k) + P_m^{(13)}(k)\,,
\eea
with
\begin{align}
P_m^{(22)}(k) &= 2  \int_\bq \left[F_2(\bq,\bk-\bq)\right]^2 P_0(q)P_0(|\bk-\bq|)\,, \\
P_m^{(13)}(k) &= 6  P_0(k)\int_\bq F_3(\bq,-\bq,\bk) P_0(q)\,.
\end{align}
Here, $\int_\bq \equiv d^3q$. The symmetrized second-order kernel is given by 
\be
F_2(\bq,\bk-\bq) = \frac{k^2(7\bk.\bq+3q^2) - 10(\bk.\bq)^2}{14 q^2 |\bk-\bq|^2},
\ee
while the symmetrized third-order kernel is given by 
\begin{align}
F_3(\bq,-\bq, \bk) &= \frac{1}{|\bk-\bq|^2} \left[ \frac{5k^2}{63} - \frac{11 \bk.\bq}{54} - \frac{k^2 (\bk.\bq)^2}{6q^4} + \frac{19(\bk.\bq)^3}{63q^4} \right. \nonumber \\ 
&\left. -\frac{23k^2\bk.\bq}{378q^2} - \frac{23(\bk.\bq)^2}{378q^2}+ \frac{(\bk.\bq)^3}{9k^2q^2}\right].
\end{align}
The other loop contributions in Eq.  \eqref{eq:ph_1loop}, all vanishing in the limit $k\to 0$, are given by:
\begin{align}
&\cI_{\delta^2}(k)=2\int_\bq F_2(\bq,\bk-\bq)P_0(|\bk-\bq|)P_0(q),  \\
&\cI_{\mathcal{G}_2}(k)=2\int_\bq S^2(\bq,\bk-\bq)F_2(\bq,\bk-\bq)P_0(|\bk-\bq|)P_0(q),   \\
&\cI_{\delta^2\delta^2}(k)=2\int_\bq \left[P_0(|\bk-\bq|)P_0(q) - P_0^2(q)\right],  \\
&\cI_{\mathcal{G}_2\mathcal{G}_2}(k)=2\int_\bq \left[S^2(\bq,\bk-\bq)\right]^2 P_0(|\bk-\bq|)P_0(q),  \\
&\cI_{\delta^2\mathcal{G}_2}(k)=2\int_\bq S^2(\bq,\bk-\bq)P_0(|\bk-\bq|)P_0(q),  \\
&\cF_{\mathcal{G}_2}(k)=4P_0(k)\int_\bq S^2(\bq,\bk-\bq)F_2(\bq,-\bk)P_0(q),
\end{align}
where the kernel $S^2$ is the Fourier transform of the Galileon operator and can be written as:
\be
S^2(\bk_1, \bk_2) = \left(\frac{\bk_1.\bk_2}{k_1k_2}\right)^2 -1\,.   
\ee

\section{Theoretical Halo Biases} \label{app:halo_biases}

We use the Sheth-Torman predictions of linear and quadratic local-in-matter bias parameters \cite{Sheth:1999mn,Scoccimarro:2000gm}, 
\begin{align}
    b_1(M) &= 1 + \epsilon_1 + E_1, \\
    b_2(M) &= 2\left(1 - \frac{17}{21}\right)(\epsilon_1 + E_1) + \epsilon_2 + E_2, 
\end{align}
where 
\begin{align}
\epsilon_1 = \frac{\alpha \nu^2 - 1}{\delta_c}, &\quad \epsilon_2 = \frac{\alpha \nu^2}{\delta_c^2}(\alpha \nu^2 -3),  \\
E_1 = \frac{2 p/\delta_c}{1+(\alpha\nu^2)^p}, &\quad E_2 = E_1\left(\frac{1+2p}{\delta_c}+2\epsilon_1\right),
\end{align}
with $\alpha = 0.707$ and $p = 0.3$. We use the co-evolution predictions (see e.g., \cite{Chan:2012jj,Baldauf:2012hs}) to relate $b_{\cG_2}$ and $b_{\Gamma_3}$ to linear bias $b_1$,
\begin{align}
b_{\cG_2}(M) &= -\frac{2}{7}\left[b_1(M)-1\right], \qquad b_{\Gamma_3}(M) = \frac{23}{42} \left[b_1(M)-1\right].
\end{align}

\bibliographystyle{utphys}
\bibliography{LIM_PS_HM}
\end{document}